\documentclass[10pt]{iopart}
\expandafter\let\csname equation*\endcsname\relax
\expandafter\let\csname endequation*\endcsname\relax

\usepackage{amsmath}
\usepackage{graphicx}
\usepackage[x11names]{xcolor}

\usepackage{amssymb}

\usepackage{tikz}
\usetikzlibrary{shapes.geometric}
\newcommand{\tikzcircle}[2][Gold1,fill=Gold1]{\tikz[baseline=-0.5ex]\draw[#1,radius=#2] (0,0) circle ;}%
\newcommand{\sqdiamond}[1][fill=black]{\tikz [x=1.2ex,y=1.2ex,line width=.1ex,line join=round, yshift=-0.285ex] \draw  [#1]  (0,.5) -- (.5,1) -- (1,.5) -- (.5,0) -- (0,.5) -- cycle;}%
\newcommand{\MyDiamond}[1][fill=black]{\mathop{\raisebox{-0.275ex}{$\sqdiamond[#1]$}}}

\usepackage{epstopdf}
\usepackage{bm}
\usepackage[utf8]{inputenc}
\usepackage[T1]{fontenc}
\usepackage[english]{babel}

\newcommand{\ket}[1]{\left\lvert{#1}\right\rangle}
\newcommand{\bra}[1]{\left\langle{#1}\right\rvert}
\newcommand{\op}[2]{\left\lvert{#1}\middle\rangle\!\middle\langle{#2}\right\rvert}
\let\divisionsymbol\div
  
\newcommand{\vect}[1]{\boldsymbol{#1}}

\begin{document}

\title[Spin-squeezing from continuous measurements]{Analysis of spin-squeezing generation in cavity-coupled atomic ensembles with continuous measurements}

\author{A. Caprotti}
\address{Istituto Nazionale di Ricerca Metrologica, Strada delle Cacce 91, I-10135 Torino, Italy}
\address{University of Vienna, Faculty of Physics, Vienna Center for Quantum Science and Technology (VCQ), Boltzmanngasse 5, 1090 Vienna, Austria}
\author{M. Barbiero}
\address{Istituto Nazionale di Ricerca Metrologica, Strada delle Cacce 91, I-10135 Torino, Italy}
\author{M. G. Tarallo}
\ead{m.tarallo@inrim.it}
\address{Istituto Nazionale di Ricerca Metrologica, Strada delle Cacce 91, I-10135 Torino, Italy}
\author{M. G. Genoni}
\ead{marco.genoni@unimi.it}
\address{Dipartimento di Fisica ``Aldo Pontremoli'', Universit\`a degli Studi di Milano, via Celoria 16, I-20133 Milano, Italy}
\address{INFN, Sezione di Milano, I-20133 Milano, Italy}
\author{G. Bertaina}
\ead{g.bertaina@inrim.it}
\address{Istituto Nazionale di Ricerca Metrologica, Strada delle Cacce 91, I-10135 Torino, Italy}

\begin{abstract}
We analyze the generation of spin-squeezed states via coupling of three-level atoms to an optical cavity and continuous quantum measurement of the transmitted cavity field in order to monitor the evolution of the atomic ensemble. Using analytical treatment and microscopic simulations of the dynamics, we show that one can achieve significant spin squeezing, favorably scaling with the number of atoms $N$. However, contrary to some previous literature, we clarify that it is not possible to obtain Heisenberg scaling without the continuous feedback that is proposed in optimal approaches. In fact, in the adiabatic cavity removal approximation and large $N$ limit, we find the scaling behavior $N^{-2/3}$ for spin squeezing and $N^{-1/3}$ for the corresponding protocol duration. 
These results can be obtained only by considering the curvature of the Bloch sphere, since linearizing the collective spin operators tangentially to its equator yields inaccurate predictions.
With full simulations, we characterize how spin-squeezing generation depends on the system parameters and departs from the bad cavity regime, by gradually mixing with cavity-filling dynamics until metrological advantage is lost. Finally, we discuss the relevance of this spin-squeezing protocol to state-of-the-art optical clocks.
\end{abstract}

\maketitle

\noindent{\it Keywords\/} {Spin squeezing, Continuous measurements, atomic clocks, cavity quantum electrodynamics}

\section{Introduction}\label{sec:intro}

Quantum sensors based on atomic ensembles, such as atomic clocks, gyroscopes, magnetometers, etc., have nowadays reached and surpassed their classical counterparts. Their standard quantum limit due to the measurement noise (quantum projection noise~\cite{Itano_1993}) determines the optimal precision obtainable using uncorrelated atoms. It can be surpassed by a squeezing factor $\xi^2<1$, by introducing quantum correlations~\cite{PEZZE_2018}.
The simplest entangled state offering metrological gain is the spin-squeezed state (SSS)~\cite{KITAGAWA_1993,SpinSqueezingReview}. Over the past decade, SSSs have been demonstrated in several systems~\cite{PEZZE_2018}, including interacting Bose-Einstein condensates~\cite{Riedel2010,Hamley2012,Gross2012}, ions~\cite{Bohnet2016}, and neutral atomic ensembles. Among different techniques, SSSs have been produced by quantum non-demolition (QND) measurement~\cite{Kuzmich_GenerationSpinSqueezing_2000,Appel_2009,SchleierSmith2010,Bohnet_2014,Cox2016,Hosten2016a,Huang2023,Serafin_NuclearSpinSqueezing_2021}, collective spin ensembles with cavity-mediated interactions~\cite{SSMITH_2010,Braverman_2019,Li_CollectiveSpinLightLightMediated_2022}, and Rydberg coupling~\cite{Eckner_Realizingspinsqueezing_2023,Bornet_Scalablespinsqueezing_2023}. 

In neutral atoms, cavity-aided collective spin measurements enabled up to $20$~dB of metrologically useful spin squeezing~\cite{Cox2016,Hosten2016a} involving transitions in the radio frequency (RF) domain, i.e. 5-10 orders of magnitude smaller than optical frequencies where the best atomic clocks currently work~\cite{Beloy2021,Bothwell2022}. Recently, proof-of-principle experiments employing cavity-aided measurements have achieved spin squeezing on an optical transition~\cite{PedrozoPenafiel2020,Robinson_Directcomparisontwo_2024} and improved clock performances of a state-of-the-art optical clock~\cite{Bowden2020}. 

\begin{figure}[pb]
 \centering
 \includegraphics[width=0.7\columnwidth]{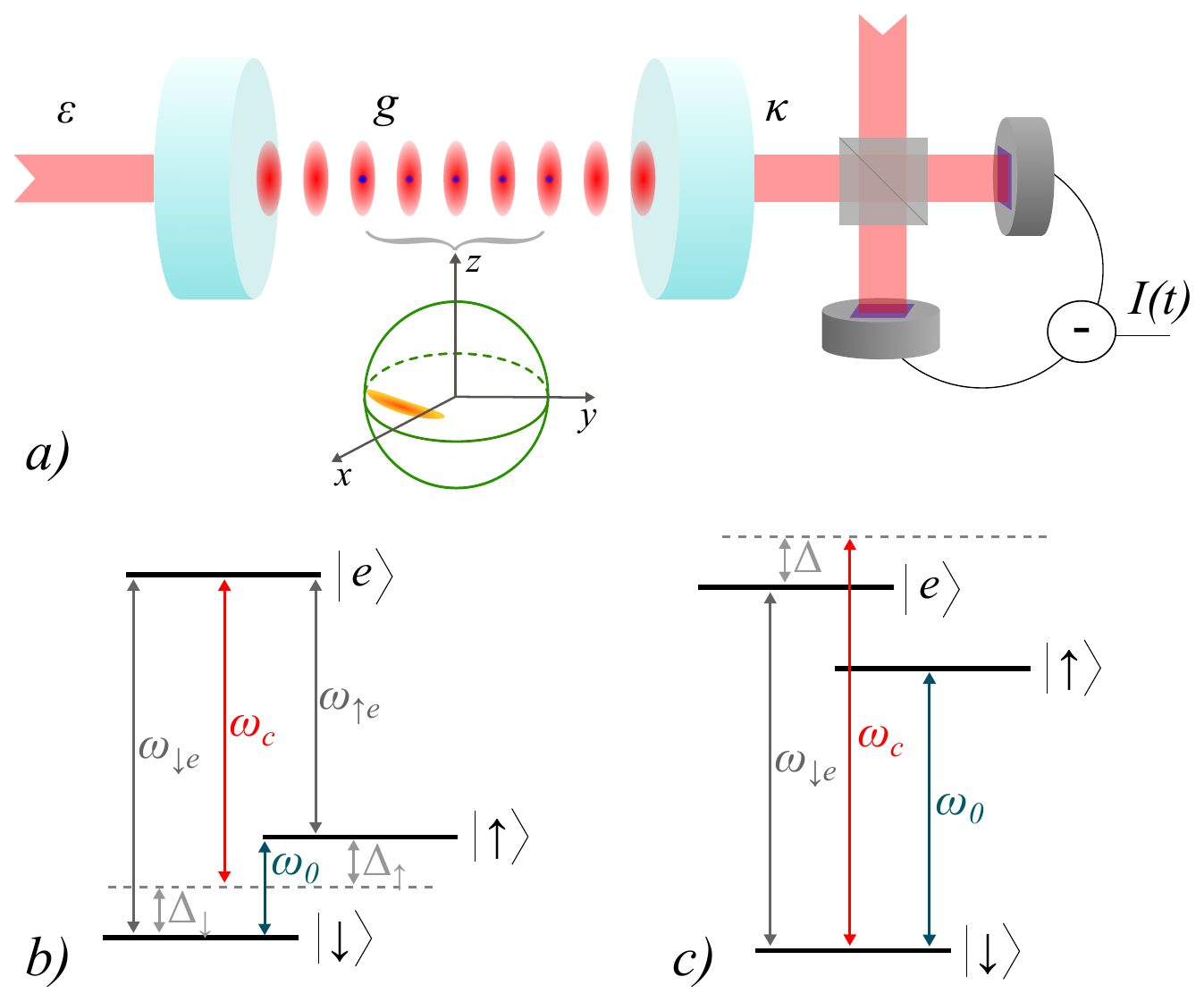}
 \caption{Panel a: Graphical scheme of the considered protocol. Three-level atoms are coupled to a cavity with coupling $g$. The cavity is driven by amplitude $\varepsilon$ and transmits at rate $\kappa$. The transmitted light is continuously measured via homodyne detection, yielding photocurrent $I(t)$. The provided information on the atomic state reduces the population difference uncertainty between the $\uparrow$ and $\downarrow$ clock states, mapping to smaller spread along the $z$ direction in the collective Bloch sphere. In the absence of continuous feedback, the average spin does not lie on the equator, but it is conditioned on the measurement result. 
Panel b: energy levels and detunings in the $\Lambda-$configuration. Panel c: $V-$configuration. $\omega_0$ and $\omega_c$ are the clock and the cavity mode frequencies, respectively.}
 \label{fig:scheme}
\end{figure}

Measurement protocols based on continuous monitoring~\cite{ALBARELLI2024129260} have been extensively studied for quantum state engineering purposes ~\cite{WisemanMilburn,JacobsBook}, leading to the outstanding experimental results observed in~\cite{Rossi2018,Magrini2021,Tebbenjohanns2021} for the cooling of a quantum mechanical oscillator towards its quantum ground state. In particular much theoretical effort has been devoted to the exploitation of this kind of protocols for the generation of metrologically useful quantum states, such as squeezed states of quantum harmonic oscillators or of spin-squeezed states for atomic ensembles ~\cite{Wiseman1993,Wiseman1994,Thomsen2002,Thomsen2002a,GEREMIA2003, MolmerMadsen2004,MADSEN_2004,Nielsen2008,SerafozziMancini,Szorkovszky2011,Genoni2013PRA,Genoni2015NJP,Hofer2015,Albarelli_Ultimatelimitsquantum_2017,Brunelli2019PRL,DiGiovanni2021,Fallani2022,Isaksen_Mechanicalcoolingsqueezing_2023,ROSSI_2020,Binefa_2021}. The physical intuition behind these protocols is the following: by continuously monitoring a particular observable of the quantum system, for example a spin operator for an atomic ensemble, the variance of such operators will decrease reaching eventually values below the so-called standard quantum limit, fixed by the fluctuations of the corresponding coherent ({\em classical}) states. The information gain from continuous monitoring is explicitly used either to perform continuous feedback on the state, resetting the target observable to a fixed value and generating deterministic, unconditional spin squeezing, or it can be stored for subsequent use. In the latter case, the final estimation for the observable of interest, namely the bias, and the generated spin squeezing are \emph{conditional}, namely they depend on the specific series of measurement outcomes.  Spin squeezing is a useful resource when created in the initial state of a Ramsey protocol: in this case, \emph{conditional} spin-squeezing is explicitly realized by exploiting the estimated bias to construct unbiased more precise estimators for the target observable, or to precisely enact a single final feedback.

Continuous monitoring of the collective spin operator of an atomic system can be achieved by engineering a dispersive coupling between the atoms and a cavity field driven by an external laser. By performing a continuous homodyne detection on the cavity output, one is indeed implementing a quantum non-demolition (QND) measurement of the spin operator~\cite{Thomsen2002,Thomsen2002a,GEREMIA2003, MolmerMadsen2004,MADSEN_2004,Nielsen2008,Serafin_NuclearSpinSqueezing_2021,Albarelli_Ultimatelimitsquantum_2017,ROSSI_2020,Binefa_2021}. In this work we will employ both analytical treatment and full cavity quantum electrodynamics simulations, to characterize how and under which assumptions this kind of interaction and consequent dynamics can be achieved in specific atomic ensembles, with a particular attention to the application on future optical clocks.

Our analysis will unveil a surprising result regarding one of the major questions when devising spin-squeezing protocols, namely the determination of the scaling exponent of the spin-squeezing parameter for large number of particles $\xi^2\propto N^{-\alpha}$. Continuous feedback protocols often reach Heisenberg scaling $\alpha=1$~\cite{Thomsen2002,Thomsen2002a}. Reduction from Heisenberg scaling in spin systems is typically due to the curvature of the collective Bloch sphere, which causes the backaction of the squeezing operation to reduce contrast. As an important example, one finds $\alpha=2/3$ for the one-axis twisting Hamiltonian (OAT)~\cite{KITAGAWA_1993}. In the literature, it is often implied that continuously monitoring protocols feature Heisenberg scaling in the $N\to\infty$ limit, even in the absence of continuous feedback. Here, we analytically and numerically show that instead a curvature effect is also present in such case. This important result impacts some qualitative findings regarding spin-squeezing scaling presented in previous works (e.g.~\cite{GEREMIA2003,Binefa_2021,Zhang_StochasticMeanfieldTheory_2023,Barberena_Tradeoffsunitarymeasurement_2023}).

Another relevant experimental parameter to be considered is the collective state preparation time, which in the case of the atom-cavity coupled system coincides with the cavity interaction time $t$. This time must be optimized in order to reduce atom-cavity scattering and decoherence, and to minimize aliasing noise due to the added dead-time in the atomic sensor~\cite{Quessada_2003}. Also in this case, we find that curvature impacts the expected result by introducing a dependence on the number of atoms.

The article is organized as follows. In Sec.~\ref{sec:Model} we introduce in detail the considered model of continuously measured cavity-coupled atoms, and describe the master equations used to study its dynamics. In Sec.~\ref{sec:results} we discuss the analytical treatment of the cavity-removal regime and the results of our full simulations, concerning the optimal spin squeezing, the time at which this is expected, and their scaling with atom number, which is impacted by the interplay between the absence of continuous feedback, Bloch sphere curvature, and atom-cavity coupling. In Sec.~\ref{sec:SrRel} we discuss the relevance of our results for optical clocks, and in Sec.~\ref{sec:conclusions} we draw our conclusions. The Appendices detail the adiabatic elimination of the atomic excited state, the tangential spin-squeezing parameter evaluation, the analytical derivations, and the computational details concerning our simulations.

\section{Model and methods}\label{sec:Model}

The considered system, as schematically depicted in figure~\ref{fig:scheme}, is the simplest model of a cavity-enhanced atomic optical clock: an ensemble of $N$ three-level uncorrelated atoms placed in a driven-dissipative optical cavity, which mediates an effective interaction between them. We assume that a deep optical lattice freezes the translational degrees of freedom of the atoms (Lamb-Dicke regime), so that only the internal states are relevant. The interaction between an atomic ensemble and a light mode in a high-finesse optical cavity has been intensively studied for the generation of both atom-light and atom-atom entanglement~\cite{Kloc_2017,Li_CollectiveSpinLightLightMediated_2022}. We focus on the generation of the input (spin-squeezed) collective state of a Ramsey protocol, deferring to future work the analysis of the entire preparation/interrogation cycle, including the role of clock laser noise and dead time in a closed-loop optical clock~\cite{Schulte_Prospectschallengessqueezingenhanced_2020,Braverman_Impactnonunitaryspin_2018}.

Throughout the paper we set $\hbar=1$, meaning that we measure energy in units of angular frequency. The clock states are labeled $\downarrow, \uparrow$, and the clock frequency is $\omega_0$. 
For the clock states subspace, we use the standard pseudo-spin-$1/2$ representation: 
$\hat{s}_x = (\op{\uparrow}{\downarrow} + \op{\downarrow}{\uparrow})/2$, $\hat{s}_y = i( \op{\downarrow}{\uparrow} - \op{\uparrow}{\downarrow})/2$, $\hat{s}_z =(\op{\uparrow}{\uparrow} - \op{\downarrow}{\downarrow})/2$, obeying the algebra $[\hat{s}_j, \hat{s}_k]=i\epsilon_{jkl}\hat{s}_l$. 
The global atomic ensemble is characterized by a collective spin vector $\hat{\vect{J}} = \sum^{N}_i \hat{\vect{s}}^{(i)}$, and $\hat{J}_z = (\hat{N}_{\uparrow} - \hat{N}_{\downarrow})/2$ corresponds in particular to the difference of population of the two clock states.

We initially focus on the $\Lambda-$level configuration (figure~\ref{fig:scheme}b) interacting with a single cavity mode $\hat{c}$ with balanced couplings $g_\uparrow=g_\downarrow\equiv g$ and symmetric cavity detunings $\Delta_\uparrow=-\Delta_\downarrow\equiv\Delta=\omega_0/2$. We thus consider the quantized Stark-shift Hamiltonian
\begin{equation}\label{eq:ham_eff}
 \hat{H}_{a} = \sum_i \frac{g^2}{\Delta} \hat{c}^\dagger\hat{c} \left(\ket{\uparrow}_i \bra{\uparrow} - \ket{\downarrow}_i \bra{\downarrow}\right) = \frac{2 g^2}{\Delta} \hat{n}\, \hat{J}_z
\end{equation}
in the rotating frame of the bare atomic levels and cavity mode, whose derivation from the cavity-coupled three-level Hamiltonian is reported in~\ref{app:elimination}. $\hat{n}=\hat{c}^\dagger\hat{c}$ is the cavity photon number operator.
Having removed the atomic auxiliary excited state $e$, the population difference of the clock states remains constant, as the operator $\hat{J}_z$ commutes with the effective Hamiltonian. This dispersive interaction thus provides a means of QND measurement of $\hat{J}_z$. 

The fundamental request to perform the above excited-state adiabatic elimination is for the detuning (and thus the clock frequency $\omega_0$) to be much larger than any other frequency, so that the transitions to and from the excited state happen on a much smaller time-scale than any other process. 
This also translates into a request regarding the cavity dynamics: from the point of view of the atomic ensemble, the interaction factor $2 g^2 \hat{n}/\Delta$ corresponds to a frequency shift, which must, for consistency, be much smaller than $\Delta$. This corresponds to the request that the average number of photons is 
\begin{equation}\label{eq:erem_cond}
 \langle\hat{n}\rangle \ll \left(\Delta/g\right)^2
\end{equation}

Up to now we described the dynamics of the atomic component and its interaction with the cavity mode. The internal dynamics of the cavity is given by a usual single mode bosonic Hamiltonian (neglecting zero-point energy) and an additional driving term, which, in the laboratory reference frame, is given by 
$\hat{H}_{c}^{\text{lab}} = \omega_c \hat{n} +
 \varepsilon\left(\hat{c}\,e^{i\omega_D t} + \hat{c}^\dagger e^{-i\omega_D t}\right)$, where $\omega_D$ is the driving laser frequency and $\varepsilon$ is the driving amplitude.
We consider a loss term characterized by a transmission rate $\kappa$ corresponding to photon decay to the external environment through the cavity walls. Driving amplitude and transmission rate are not independent, but related by $\varepsilon=\sqrt{\kappa P/\omega_D}$, where $P$ is the experimentally widely tunable pumping power. When working in the cavity frame of reference, the cavity Hamiltonian is
 $\hat{H}_{c} = 
 \varepsilon\left(\hat{c}\,e^{-i\delta_D t}+ \hat{c}^\dagger e^{i\delta_D t}\right)$, 
where $\delta_D = \omega_c -\omega_D$ is the detuning between the cavity mode and the driving laser. In this work, we focus on the case of resonant driving laser $\delta_D=0$, where there is no explicit time dependence. This regime enhances the feasibility of measurement-induced spin-squeezing generation, while the nearly-detuned regime has been also considered for a deterministic generation of induced-interaction squeezing, which has been often dubbed "coherent cavity feedback"~\cite{SSMITH_2010,Li_CollectiveSpinLightLightMediated_2022} (not to be confused with the feedback used in some continuous measurement protocols).
In the absence of coupling to the atomic transitions, the number of photons which occupy the cavity in the steady-state at large times would stabilize at
\begin{equation}\label{eq:num_photons}
	n_0 = \left(\frac{2\varepsilon}{\kappa}\right)^2= \frac{4 P}{\kappa \omega_D}.
\end{equation}
Therefore the total Hamiltonian of the atom-cavity system that here we consider is $\hat{H} = \hat{H}_{a} + \hat{H}_{c}$.

The main figure of merit of the considered protocol is the spin-squeezing parameter. In a general sense, squeezed states have reduced variance for a certain observable, at the cost of increased variance for a non-commuting observable~\cite{SpinSqueezingReview}. Following the definition by Kitagawa and Ueda~\cite{KITAGAWA_1993}, $N$ two-level atoms being described by a collective spin with maximum magnitude $J = N/2$ are in a spin-squeezed state (SSS) if the variance of one spin component $\hat{J}_\perp$, normal to the mean spin vector $\langle\hat{\vect{J}}\rangle$, is smaller than the variance of a coherent spin state (CSS), $\Delta^2 \hat{J}_\perp < J/2$.
To be metrologically relevant, such variance is weighted by the contrast $\mathcal{C}=|\langle\hat{\vect{J}}\rangle|^2/J^2$, yielding Wineland's spin-squeezing parameter~\cite{WINELAND_1992,Wineland_Squeezedatomicstates_1994}:
\begin{equation}\label{eq:squeezing_xi_def}
 \xi^2 = \min_\perp \left(\frac{\Delta^2\hat{J}_\perp}{J\mathcal{C}/2}\right)
\end{equation}
The spin-squeezing parameter of a CSS is $\xi^2_{\text{CSS}}= 1$, corresponding to the standard quantum limit (SQL). This represents the best scaling available using uncorrelated atoms. Metrologically useful spin squeezing corresponds to $\xi^2 < 1$.

\subsection{Continuous measurement dynamics}\label{subsec:continuousMeasurements}

The main idea in the scheme that we analyze to generate spin squeezing is that, since the dynamics described by the Hamiltonian~\eqref{eq:ham_eff} couples directly the collective spin $z-$component to the bosonic field, we may obtain information on that particular observable from measurements on the photonic degrees of freedom, without having to directly perturb the atomic ensemble.
A QND measurement does not in fact perform a destructive projective measurement on the system itself, but instead acts on the environment coupled to the considered system~\cite{Braginsky1980,Braginsky_Quantumnondemolitionmeasurements_1996,Clerk_Introductionquantumnoise_2010}.
In particular, through continuous homodyne sensing of the transmitted photonic field $\sqrt{\kappa}\hat{c}$~\cite{GARDINER_1985}, one can detect the phase shift proportional to the atomic population difference, thus obtaining information regarding $\hat{J}_z$~\cite{Thomsen2002,Thomsen2002a,MADSEN_2004,Albarelli_Ultimatelimitsquantum_2017}. 

The dynamics of the internal system is described by a stochastic master equation (SME) for the density matrix conditioned on the measurement outcome $\hat{\rho}_c$, which contains a decoherence term, due to the interaction with the external environment, and a stochastic term which instead describes the non-linear evolution of the system due to the performed measurement~\cite{WisemanMilburn,ALBARELLI2024129260}:
\begin{align}\label{eq:master_equation}
 &d\hat{\rho}_c &=\phantom{=}& -i\left[\hat{H},\hat{\rho}_c\right] dt   + \kappa \;\mathcal{D}[\hat{c}] \;\hat{\rho}_c \,dt 
 +\sqrt{\eta\kappa} \, \mathcal{H}[\hat{c}e^{-i\varphi}]\;\hat{\rho}_c \,dW_t 
 \nonumber\\
 &I(t) dt &=\phantom{=}& \sqrt{\eta \kappa}\langle \hat{c}e^{-i\varphi}+\hat{c}^\dagger e^{i\varphi} \rangle_c \,dt + dW_t 
\end{align}
where the notation $\langle \hat{A}\rangle_c$ indicates the expectation value of operator $\hat{A}$ with the conditional density matrix $\hat{\rho}_c$ and we have introduced the Lindbladian superoperator $\mathcal{D}[A]\bullet = A\bullet A^\dagger - \frac{1}{2}\lbrace A^\dagger A,\bullet\rbrace$, and the non-linear superoperator $\mathcal{H}[A]\bullet = A\bullet+\bullet A^\dagger - Tr\left[\bullet( A + A^\dagger)\right]\bullet$.
The so-called photocurrent $I(t)$ is the outcome of normalized homodyne detection at each time step, and the parameter $\varphi$ represents the phase of the local oscillator to which the photons exiting the cavity are coupled in order to perform the sensing of the bosonic field~\cite{leonhardt1997measuring}. In particular, we choose $\varphi = 0$, which corresponds to a measurement of the field quadrature $(\hat{c}+\hat{c}^\dagger)/\sqrt{2}= \hat{x}$ in the standard basis. The photocurrent is biased by system observables, but its quantum measurement noise is described by the Wiener increment $dW_t$. By definition of the second of equations~\eqref{eq:master_equation}, the Wiener increment can be found as the difference between the actual measured photocurrent $I(t)$ and its expected value at each time step. Experimentally, it acts as an innovation increment, since including it in the first of equations~\eqref{eq:master_equation} allows for considering information about the system that would normally be lost to the environment. 
When mathematically modeling an ensemble of possible experiments, as we do, the Wiener increment is a stochastic variable following a Gaussian distribution with mean $\mathrm{E}[dW_t]=0$ and variance $\mathrm{E}[dW^2_t ] = dt$. Its characteristic property is that, in the infinitesimal time-step limit, its square is not random but deterministically $dW^2_t \equiv dt$. 

The parameter $0\leq \eta\leq 1$ phenomenologically accounts for measurement efficiency. Optimally, for $\eta=1$, all photons leaking from the cavity would undergo successful homodyne interference and detection, while in the opposite case of null efficiency, equation~\eqref{eq:master_equation} would reduce to a Lindblad master equation where the only effect of cavity transmission is to introduce dissipation. 

It is important to realize that conditional evolution implies that the measurement outcome time series must be put to effective use in order for it to correspond to an information gain to be exploited in a subsequent Ramsey protocol. One possibility is to perform continuous feedback conditioned on measurement outcomes, whose description, however, requires additional terms in~\eqref{eq:master_equation} (see~\cite{Thomsen2002,Thomsen2002a}). If continuous feedback is not enacted, one should track $\langle\hat{J}_z\rangle_c$ via all the measurement outcomes, obtaining a best estimate at the end of the spin-generation protocol. Then, two possibilities are equivalently described by~\eqref{eq:master_equation}: either acting on the system only at the end of the QND protocol with a single feedback depending on the inferred $\langle\hat{J}_z\rangle_c$, corresponding to a single final unitary rotation of the spin-squeezed state towards the equator of the Bloch sphere (see figure~\ref{fig:scheme}a); or employing the same inferred $\langle\hat{J}_z\rangle_c$ as a bias to be subtracted from the projective measurement of $\hat{J}_z$ at the end of Ramsey protocol~\cite{SchleierSmith2010}, which has the advantage of not introducing further experimental operations. Notice that a single final feedback is not equivalent to performing continuous feedback, and it is still described by our treatment. To this aim, we now discuss how $\langle\hat{J}_z\rangle$ can be experimentally inferred.
The master equation defined in~\eqref{eq:master_equation} can be used to determine the evolution of the expectation value of relevant quantities, for example $\langle\hat{J}_z\rangle_c$.
From the definition of expectation value as $\langle\hat{J}_z\rangle = \Tr[\hat{\rho}\;\hat{J}_z]$, we get the conditional evolution equation: 
\begin{align}
 d\langle \,\hat{J}_z\,\rangle_c 
 &= 
 \sqrt{2\eta\kappa}\,\left[ \langle \hat{J}_z\hat{x}\rangle_c
 - \langle \, \hat{J}_z\rangle_c \langle \hat{x}\rangle_c \right] \,dW_t \,
 \label{eq:jz_master_equation}
 \nonumber\\ 
 I(t)\;dt &= \sqrt{2\eta\kappa}\langle \hat{x} \rangle_c \,dt + dW_t \;.
\end{align}

Since $\hat{J}_z$ commutes with the Hamiltonian, as expected the evolution of its average value is determined only by the stochastic increment that depends on the measurement outcome.
At first it may seem that the evolution of the expectation value, and thus of the spin-squeezing parameter, may be obtained solely from the photocurrent measurements and the evolution of the measurement quadrature.
However, even though the state density matrix does not appear directly in~\eqref{eq:jz_master_equation}, it is still necessary to determine the conditional increment.
At any given time, it is thus necessary to know the full conditional density matrix in order to determine the value of this random increment, and it is not possible to determine exactly the conditional evolution of the expectation value of relevant observables without also knowing the conditional trajectory of the full state. However, as we also remark in the following, it is possible in certain scenarios to approximately determine its value from the photocurrent and also cancel this stochastic contribution via real-time feedback. 

Finally, we mention that in our model~\eqref{eq:master_equation}, we do not consider other decoherence sources, like atomic decay, assuming that their time scales are long, when compared to the relevant dynamics. However, we will briefly discuss them in Sec.~\ref{sec:SrRel}.

\subsection{Adiabatic cavity removal}\label{subsec:cavityRemoval}

As shown in equation~\eqref{eq:ham_eff}, the cavity interacts with the atomic ensemble with a maximum absolute frequency shift
\begin{equation}\label{eq:eff_frequency}
  \delta\omega \equiv \frac{g^2}{\Delta} N \geq \frac{2 g^2}{\Delta} \left|\langle\,\hat{J}_z\,\rangle \right|
\end{equation}
The other process in which the cavity photons are involved is the cavity loss, which happens at a rate $\kappa$, and corresponds to the information acquisition rate, when $\eta=1$. 
When this rate is much larger than the effective shift per photon, 
\begin{equation}\label{eq:bc_condition}
 \kappa \gg \delta\omega
\end{equation}
the system is said to be in the so-called "bad cavity regime", where the information on the atoms encoded in the photon leaving the cavity is transferred directly to the detector (when efficiency is maximal), as if the measurements were performed directly on the spin system. 
The optical cavity thus represents a "medium" through which information is transferred and, much like the excited state in~\eqref{eq:ham_eff}, it can be adiabatically removed~\cite{DOHERTY_1999}.
Following the same scheme, one obtain the effective dynamics described by the following SME:
\begin{align}\label{eq:master_equation_cavrem}
 d\hat{\rho}_c &= \tilde{\kappa} \;\mathcal{D}[\hat{J}_z] \;\hat{\rho}_c \,dt + \sqrt{\eta\tilde{\kappa}} \, \mathcal{H}[\hat{J}_z]\;\hat{\rho}_c \, dW_t
 \nonumber\\ 
 I(t)\;dt &= 2 \sqrt{\eta\tilde{\kappa}}\langle \hat{J}_z \rangle_c \, dt + dW_t
\end{align}
where the density matrix now refers only to the atomic Hilbert space, and the effective transmission rate is~\cite{Thomsen2002}:
\begin{equation}\label{eq:k_eff}
 \tilde{\kappa} = 4 \left(\frac{2g^2}{\Delta}\right)^2 \frac{ n_0}{\kappa}\;.
\end{equation}
Here, since the photons are not dynamical anymore and their frequency shift is negligible, we have assumed that their number is equal to the stationary one in noninteracting cavity, equation~\eqref{eq:num_photons}.
We notice that there is no longer any Hamiltonian term, apart from a constant Stark shift that has been included in the reference frame: the cavity-atom interaction is directly embodied by the dissipative and measurement terms of the effective SME. The generation of spin squeezing under this evolution has been investigated in great detail in~\cite{Thomsen2002,Thomsen2002a}. As for equations~\eqref{eq:jz_master_equation}, also in this case one observes a stochastic evolution for $\langle \hat{J}_z\rangle$, given by the equation $d\langle\hat{J}_z\rangle_c=2\sqrt{\eta\tilde{\kappa}}\;(\Delta^2\hat{J}_z)_c\; dW_t$. This may be corrected exactly via Markovian feedback by solving the full trajectory of the conditional state; the corresponding feedback scheme leads to an unconditional Heisenberg-limited spin squeezing. It was also shown that an approximate feedback, depending only on the photocurrent results and not on the full trajectory, allows for deriving the following approximate analytical solution valid for short-to-intermediate times:
\begin{equation}\label{eq:fb_squeezing_sol}
 \xi^2_\text{F}=\dfrac{e^{\tilde{\kappa}t}}{1+\eta N\tilde{\kappa}t} 
\end{equation}
from which one obtains a minimum spin-squeezing parameter following Heisenberg scaling:
\begin{equation}\label{eq:fb_squeezing_min}
 \xi^2_\text{F,m}=e/(\eta N)\,,
\end{equation}
reached at the optimal time 
\begin{equation}\label{eq:fb_t_opt}
 t_\text{F,m}=1/\tilde{\kappa}\;.
\end{equation}

We now focus on the assumptions needed to perform the cavity adiabatic removal: first of all, the cavity is assumed to be in the stationary regime, so that photons follow no dynamics other than the decay into free space; given a weak interaction with the atomic ensemble, this request relates their number only on the parameters $\varepsilon$ and $\kappa$, as in~\eqref{eq:num_photons}.
Secondly, the "bad cavity" requirement that $\kappa$ must be the highest frequency (besides $\Delta$) imposes a further condition besides~\eqref{eq:bc_condition}, namely that $g^2 \langle\hat{n}\rangle/\Delta \ll \kappa$. This provides a tighter bound on the maximum average number of photons expected in the cavity than the one expressed in~\eqref{eq:erem_cond}:
\begin{equation}\label{eq:cavrem_ph_cond}
 \langle\hat{n}\rangle \ll \kappa\Delta/g^2 \ll (\Delta/g)^2.
\end{equation}

\subsection{Simulation of system dynamics}\label{subsec:simulation} 

We solve the SMEs~\eqref{eq:master_equation} and \eqref{eq:master_equation_cavrem} using the QuTiP library~\cite{QUTIP_2012, QUTIP_2013}. A very relevant speed-up is obtained by considering only the atomic Dicke sector with maximum eigenvalue $J(J+1)$ of $\hat{J}^2$, with $J=N/2$~\cite{ARECCHI_1972}. We are allowed to do so because we do not consider atomic depolarization and we choose an initial pure state in this subspace, namely a spin-coherent state with $J=N/2$. In the cavity-removal approximation, we will analytically demonstrate that measurement efficiency $\eta<1$ only introduces a constant prefactor in the minimal spin-squeezing parameter. In all our simulations we considered instead the simpler case $\eta=1$. In the case of equation~\eqref{eq:master_equation}, the initial atomic state is in a tensor product with an empty cavity. Since we focus on unit efficiency, this allows us to reduce our simulations to the corresponding stochastic Schrödinger equations (SSE)~\cite{WisemanMilburn}, with tremendous reduction of memory and computational requirements, since vectors in Hilbert space are evolved, instead of density operators. 
See~\ref{app:simulation} for details on the simulation setup.

The metrological spin-squeezing parameter of equation~\eqref{eq:squeezing_xi_def} is not simply proportional to the spin variance along $z$, but it is estimated at any given time by determining the minimal variance of the collective spin components which are perpendicular to the instantaneous mean spin vector. This corresponds to the smallest eigenvalue (normalized by the contrast) of the covariance matrix:
\begin{equation}\label{eq:cov_mat_def}
 \text{cov}_{ij}(\hat{\vect{J}}) 
 = \frac{1}{2}\langle \hat{J}_i\hat{J}_j+\hat{J}_j\hat{J}_i\rangle 
 - \langle \hat{J}_i\rangle \langle \hat{J}_j\rangle
\end{equation}
where $i,j\in\{1,2\}$ and $\hat{J}_i=\hat{\vect{J}}\cdot \vect{n}_i$, with $\vect{n}_i\perp \langle\hat{\vect{J}}\rangle_c$. The polar and azimuthal angles are defined by
 $\langle \hat{\vect{J}}\rangle_c$ by
\begin{equation}\label{eq:thetaphi}
 \cos{\theta}=\langle\hat{J}_z\rangle_c/|\langle \hat{\vect{J}}\rangle_c|\;,\qquad
 \tan{\phi} = \langle\hat{J}_y\rangle_c/\langle\hat{J}_x\rangle_c\;.
\end{equation}
Details on this evaluation can be found in~\ref{app:rotation}.

\section{Results}\label{sec:results}

\subsection{Analytical results in the cavity-removal approximation}\label{sec:analytical}

In this section, we determine an analytical expression for the conditional and average spin-squeezing parameter in the cavity-removal approximation, by analyzing the time evolution of the conditional mean spin and spin covariance. Here, we outline the derivation, while details are reported in~\ref{app:analytical}.

\begin{figure}[tb]
 \centering
 \includegraphics[width=0.7\columnwidth]{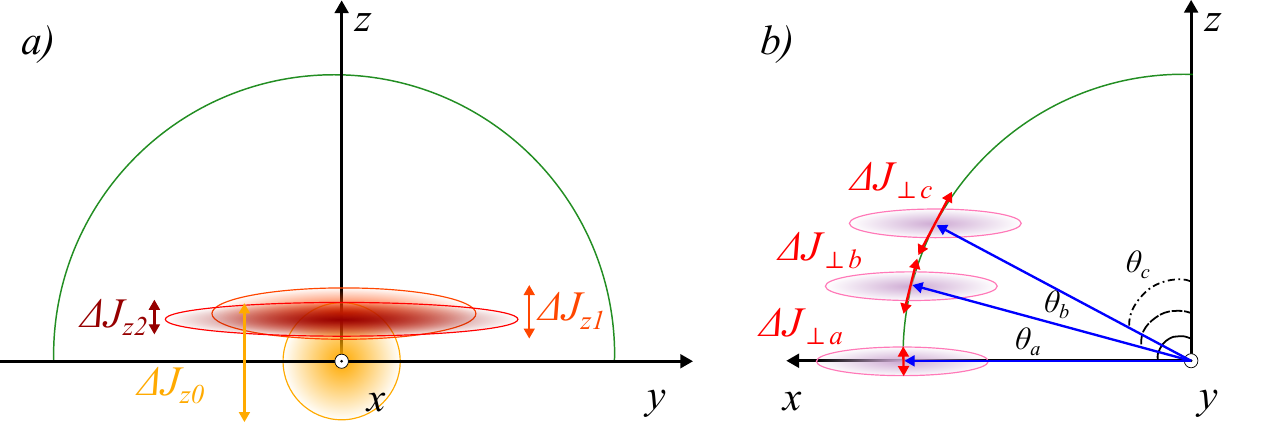}
 \caption{
 Sections of the spin covariance ellipsoid and of the collective Bloch sphere on the $y-z$ and $x-z$ planes co-rotating with the average spin. Panel a exemplifies how, along a single quantum trajectory taken at three subsequent times $0,1,2$, $\langle\hat{J}_z\rangle_c$ evolves stochastically but gradually converging, due to the $z$ variance unconditionally decreasing, while the $y$ uncertainty correspondingly increases. On panel b, 
 we exemplify the tangential-spin standard deviation $\Delta J_\perp$ for different polar angles $\theta$ of the mean spin, showing that, at fixed $z$ uncertainty, the smaller the polar angle, the higher the $x$ relative contribution is. Sizes and angles are exaggerated and reduction of contrast is not showed for clarity.}
 \label{fig:BlochSphereSections}
\end{figure}

We introduce the scaled time $\tau\equiv\tilde{\kappa} t$ and recall that the evolution of the conditional expectation value of an observable $\hat{A}$ is determined by the SME~\eqref{eq:master_equation_cavrem} via $d\langle\hat{A}\rangle_c=\Tr(\hat{A}\; d\hat{\rho}_c)$. Since the off-diagonal covariances are zero for $\tau=0$, we assume that they remain negligible along the dynamics. This is equivalent to a third order cumulant truncation, namely a Gaussian approximation, and will be confirmed by the simulations. Inspection of the SMEs for $\hat{J}_x$, $\hat{J}_y$, $\hat{J}_x^2$ and $\hat{J}_y^2$ shows then that their evolution is purely dissipative and unconditional. By taking into account the initial condition of a CSS along the positive $x$ axis, we obtain:
\begin{align}\label{eq:evolvexy}
 \langle\hat{J}_x(\tau)\rangle &= J e^{-\tau/2} \nonumber\\
 \langle\hat{J}_y(\tau)\rangle &= 0 \nonumber\\
 \langle\hat{J}_x^2(\tau)\rangle &= \frac{J^2}{2}\left( 1+e^{-2\tau}\right)+ \frac{J}{4}\left( 1-e^{-2\tau}\right) \nonumber\\
 \langle\hat{J}_y^2(\tau)\rangle &= \frac{J^2}{2}\left( 1-e^{-2\tau}\right)+ \frac{J}{4}\left( 1+e^{-2\tau}\right)
\end{align}
resulting in the following closed expressions:
\begin{align}
 \frac{\Delta^2\hat{J}_x(\tau)}{J/2} &= J\left( 1-e^{-\tau}\right)^2+ \frac{1}{2}\left( 1-e^{-2\tau}\right) \label{eq:deltaX}\\
 \frac{\Delta^2\hat{J}_y(\tau)}{J/2} &= J\left( 1-e^{-2\tau}\right)+ \frac{1}{2}\left( 1+e^{-2\tau}\right)\;.\label{eq:deltaY}
\end{align}

Conversely, it is clear that all powers of $\hat{J}_z$ evolve only via the stochastic term, due to them commuting with the dissipator. 
However, the evolution of $\Delta^2\hat{J}_z$ contains a stochastic term corresponding to the third order cumulant $\langle\hat{J}_z^3\rangle_C$ that we approximate to zero, and an additional unconditional term stemming from It\^{o} calculus, resulting in
\begin{equation}
 d\Delta^2\hat{J}_z=-4\eta\;(\Delta^2\hat{J}_z)^2 \;d\tau\;,
\end{equation}
whose solution is 
\begin{equation}\label{eq:deltaZ}
 \frac{\Delta^2\hat{J}_z}{J/2}= \frac{1}{1+2J\eta\tau}\;,
\end{equation}
This result saturates the uncertainty bound in the $y-z$ plane, for moderate times and $\eta=1$: $\Delta^2\hat{J}_z\Delta^2\hat{J}_y\ge |\langle \hat{J}_x\rangle|^2/4$ (See panel a of figure~\ref{fig:BlochSphereSections}). For large $J$, the contrast is determined by the $x$ component, yielding $\mathcal{C}(\tau)\simeq \langle\hat{J}_x(\tau)\rangle^2/{J^2}= e^{-\tau}$. Notice that the expressions found for the $y$ and $z$ components are consistent in the limit of large $J$ with the results from the Holstein-Primakoff approximation (see~\cite{Albarelli_Ultimatelimitsquantum_2017,MolmerMadsen2004,MADSEN_2004}), which is however unable to correctly describe variations of $\langle\hat{J}_x\rangle$ and $\Delta^2 \hat{J}_x$: crucially, in our case these observables should not be fixed to their initial values, $J$ and $0$, respectively, as we demonstrate in the following.

We are interested in the tangential spin-squeezing parameter, corresponding to
\begin{equation}\label{eq:tangential-squeezing}
 \xi^2(\tau,\cos^2\theta)=\frac{\Delta^2\hat{J}_\perp(\tau)}{J\mathcal{C}(\tau)/2}=
 \frac{\Delta^2\hat{J}_z(\tau)\sin^2\theta+\Delta^2\hat{J}_x(\tau)\cos^2\theta}{J\mathcal{C(\tau)}/2}\;,
\end{equation}
where $\cos\theta$ is defined by the mean spin via~\eqref{eq:thetaphi} and we again neglected the off-diagonal $x-z$ covariance. Equation~\eqref{eq:deltaX} implies that $\Delta^2\hat{J}_x$ increases quadratically with small time. Panel b of figure~\ref{fig:BlochSphereSections} then shows that the $x$ contribution to spin squeezing may become dominant if the mean spin is far from the equator. Indeed, the absence of continuous feedback in our approach implies that, even in the $J\to\infty$ limit, $\theta$ should not simply be set equal to $\pi/2$. The reason is that the statistical distribution $\mathcal{P}$ of conditional values of $\langle\hat{J}_z\rangle_c$ is constant in time and equivalent to the initial one, which is Gaussian, in the large $J$ limit, and reads:
\begin{equation}\label{eq:Zdistribution}
	\mathcal{P}\left(\langle\hat{J}_z\rangle_c\right)=\frac{1}{[2\pi \Delta^2 \hat{J}_z(0)]^{1/2}}\exp{\left(-\frac{\langle\hat{J}_z\rangle_c^2}{2\Delta^2 \hat{J}_z(0)}\right)}\;.
\end{equation}

We can now analytically evaluate the trajectory average of the conditional spin-squeezing parameter in the absence of continuous feedback of equation~\eqref{eq:tangential-squeezing}, $\xi^2_{\text{NF}}(\tau)=\mathrm{E}[\xi^2(\tau,\langle\hat{J}_z\rangle_c^2/|\langle \hat{\vect{J}}\rangle|^2)]$, by noticing that equation~\eqref{eq:analyticalSqueezingCosTheta} depends quadratically on $\langle\hat{J}_z\rangle_c$, while the rest of the expression is unconditional in our approximations, and we obtain:
\begin{multline}\label{eq:analyticalSqueezingAverage}
 \xi^2_{\text{NF}}=\int_{-\infty}^{+\infty}\mathcal{P}(q) \xi^2\left(\tau,\frac{q^2}{|\langle \hat{\vect{J}}\rangle|^2}\right)dq 
 = \xi^2\left(\tau,\frac{\Delta^2 \hat{J}_z(0)}{J^2\mathcal{C}(\tau)}\right) \\
 =\frac{\left(1-\frac{e^\tau}{N}\right)e^\tau}{1+\eta N\tau} 
 + \frac{1}{2N}\left[N\left(e^{\tau}-1\right)^2+ e^{2\tau}-1\right]\;.
\end{multline}

We have thus found that Gaussianity implies that the average spin-squeezing parameter is the one corresponding to a trajectory where $\langle\hat{J}_z\rangle_c$ is equal to the initial standard deviation of $\hat{J}_z$.

To infer the scaling of the optimal time and minimal squeezing with the number of particles, we first keep only the dominant terms of the previous expression for $N\to\infty$, and then expand for small time, presuming that the minimum occurs for $\tau\ll 1$:
\begin{equation}
 \xi^2_{\text{NF}}(\tau)\underset{N\to\infty}{\approx}\frac{1}{\eta N\tau} + \frac{1}{2}\left(e^{\tau}-1\right)^2\underset{\tau\to 0}{\approx} \frac{1}{\eta N\tau} + \frac{\tau^2}{2}\;.
\end{equation}
It is clear here that the second term, stemming from the $\Delta^2\hat{J}_x$ contribution, causes an increase of the tangential spin-squeezing parameter, in competition with $\Delta^2\hat{J}_z$ that is decreasing. The time at which the minimum is reached is dubbed the optimal time $t_\text{m}$ and is relevant when devising an experimental protocol. In our analytical approximation, it occurs for 
\begin{equation}\label{eq:analyticalTopt}
 t_{\text{NF,m}} =\frac{\tau_{\text{NF,m}}}{\tilde\kappa} = \frac{1}{\tilde\kappa (\eta N)^{1/3}}\;,
\end{equation}
corresponding to the optimal average spin-squeezing parameter
\begin{equation}\label{eq:analyticalXi2}
 \xi^2_{\text{NF,m}}=\frac{3}{2}\frac{1}{(\eta N)^{2/3}}\;.
\end{equation}
Notice that the effect of non-optimal measurement efficiency $\eta<1$ is to simply introduce a constant factor, amounting to an effectively reduced atom number $\eta N$.

\begin{figure}[tbp]
 \centering
 \includegraphics[width=0.7\columnwidth]{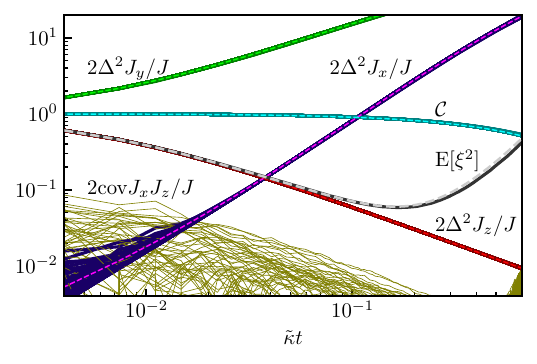}
 \caption{Comparison of the results of a cavity removal simulation with $N=160$ atoms and $\eta=1$, and the corresponding analytical expressions from Sec.~\ref{sec:analytical}.}
 \label{fig:cavity-removal}
\end{figure}

We highlight here that tangential spin-squeezing in the absence of continuous feedback is worsened by the contribution coming from an increasing variance of $\hat{J}_x$, which must be taken into account due to the curvature effect even in the $N\to\infty$ limit. This implies a minimum average spin-squeezing parameter that loses Heisenberg scaling and is on par with the OAT result in~\cite{KITAGAWA_1993}, and a corresponding optimal time which is not size independent. The found spin-squeezing exponent is slightly smaller than the one, $\alpha\simeq0.73$, recently obtained with a mean-field approach in a similar setup when also the pump detuning is optimized~\cite{Li_CollectiveSpinLightLightMediated_2022}. The scaling exponents $\alpha$, for the minimum average spin-squeezing parameter and $\beta$, for the optimal time, are consistent with the relation $\alpha+\beta = 1$, which also holds for the cavity-mediated interaction model of~\cite{SSMITH_2010,Li_CollectiveSpinLightLightMediated_2022} (see~\ref{app:exponents}). The role of Bloch sphere curvature, that we unveiled in the considered no-feedback continuous measurement protocol, and its impact on scaling, are akin to an effect discussed for a Ramsey protocol starting with an initial already spin-squeezed state~\cite{Pezze_HeisenbergLimitedNoisyAtomic_2020}, with the notable difference that here the relevant curvature is due to the initial variance of $\hat{J}_z$, while in the other case it originates from clock laser noise.

\subsection{Comparison with numerical results in the cavity-removal approximation}

\begin{figure}[tbp]
 \centering
 \includegraphics[width=0.7\columnwidth]{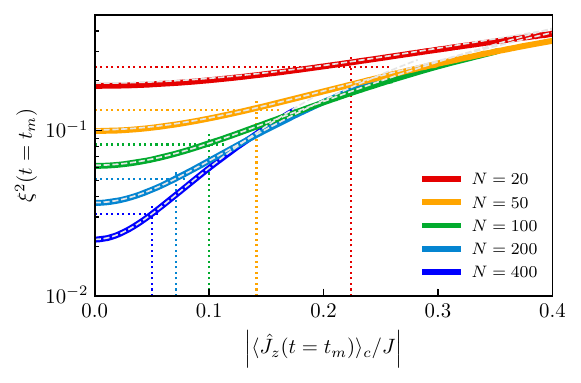}
 \caption{Correlation between $\xi^2$ and $\langle \hat{J}_z\rangle_c$ taken at the optimal time $t_\text{m}$ 
 for $M=100$ trajectories in the cavity-removal approximation with $\tilde{\kappa}=0.8\Delta$.
 Distributions are ordered by increasing atom number and horizontal dashed lines indicate the optimal average spin-squeezing parameter $\xi^2_\text{m}$.
 In order to compare results for different atomic sizes, the $z-$component of spin is normalized by the total spin $J$. Vertical lines correspond to the standard deviation of $\hat{J}_z$ in the initial state. Dashed curves are equation~\eqref{eq:analyticalSqueezingCosTheta}, where the only parameter is the chosen time from the simulations.}
 \label{fig:correlation_xi_jz}
\end{figure}

We now check the analytical results and assumptions of the previous section, via numerical solution of the SME~\eqref{eq:master_equation_cavrem}, with $\eta=1$. 

The results of a SME are crucially conditioned on the sampled noise, which is not a computational artifact, but has the physical meaning of representing a possible realization of the noise of the continuous measurements. Therefore, we both evaluate the distribution of conditional results $\langle\hat{A}(t)\rangle_{c}=\Tr [\hat{A}\hat{\rho}_{c}(t)]$ of some relevant physical quantities $\hat{A}$, and then we consider their statistical average over the $M$ trajectories $\mathrm{E}[A(t)]$. The variance of the average, which is smaller the higher $M$, is not to be confused with the variance of the distributions, which increases over time due to the wandering of the average collective spin and the absence of feedback, at odds with unconditional protocols.

In figure~\ref{fig:cavity-removal} we show the values of relevant elements of the covariance matrix, along different trajectories, together with the contrast and the average tangential spin-squeezing parameter $\mathrm{E}[\xi^2]$, and compare them to our analytical predictions (dashed lines). We simulated $N=160$ atoms and normalized the time with the effective information rate $\tilde\kappa$. The diagonal variances and the contrast are clearly almost unconditional and in excellent agreement with the results of the previous Section. Consistently, the off-diagonal covariances are essentially zero ($xy$ and $zy$ cases, not shown) or negligible ($xz$). Notice how the $z$ and $x$ variances have opposite behavior, implying that their weighted sum, the average spin-squeezing parameter, displays a minimum and then worsens. Also for this crucial quantity, the analytical expression of equation~\eqref{eq:analyticalSqueezingAverage} is in very good agreement with the numerical result in the region of the minimum (deviations at higher times stem from finite-size effects, see~\ref{app:analytical}).

In figure~\ref{fig:correlation_xi_jz} we now also check the analytical expression corresponding to~\eqref{eq:tangential-squeezing} (See equation~\eqref{eq:analyticalSqueezingCosTheta}) against simulations with varying number of atoms. At fixed time, such expression provides the correlation between the conditional value $\langle\hat{J}_z\rangle_c$ and the corresponding conditional spin-squeezing parameter. For each simulation we select the optimal time $t_\text{m}$ and plot~\eqref{eq:analyticalSqueezingCosTheta}: the agreement is again very good and discrepancies start to be noticeable only for very large $\langle\hat{J}_z\rangle_c$, where the conditional contrast should take into account not only the unconditional $x$ component, but also the $z$ contribution. Notice also that the average squeezing (horizontal lines) is consistent with the conditional squeezing corresponding to $\langle\hat{J}_z\rangle_c$ equal to the initial standard deviation (vertical lines), an agreement which increases the larger the particle number. This figure manifests the effect of Bloch sphere's curvature: trajectories which keep $\langle \hat{J}_z\rangle_c\approx 0$ display the best conditional spin squeezing, because the fixed squeezing direction $z$ is almost perpendicular to the average spin. In a continuous feedback scheme, the state is constantly realigned with the equator, thus always obtaining the maximum possible squeezing. 
On the other hand, in the absence of feedback, many trajectories will result in $\langle \hat{J}_z\rangle_c$ far from the equator, corresponding to worse spin-squeezing parameter, due to the squeezing operation not acting perpendicularly to the average spin. 
Although one might assume that the Holstein-Primakoff approximation for the collective spin is valid in the $N\to\infty$ limit, implying that the plane tangent to the Bloch sphere is always perpendicular to the equator and thus feedback is not necessary to achieve Heisenberg scaling, we have in fact analytically and numerically demonstrated that the role of curvature persists in such limit, resulting in $\alpha=2/3$ for metrological spin squeezing. Heisenberg scaling characterizes spin squeezing only if evaluated along the fixed $z$ axis~\cite{Zhang_StochasticMeanfieldTheory_2023}, which is not, however, metrologically exploitable. We numerically check the predictions concerning the scaling of the spin-squeezing parameter and the optimal time in the following Section, together with the results from the full simulations.

\begin{figure*}[bt]
 \centering
 \includegraphics[width=\textwidth]{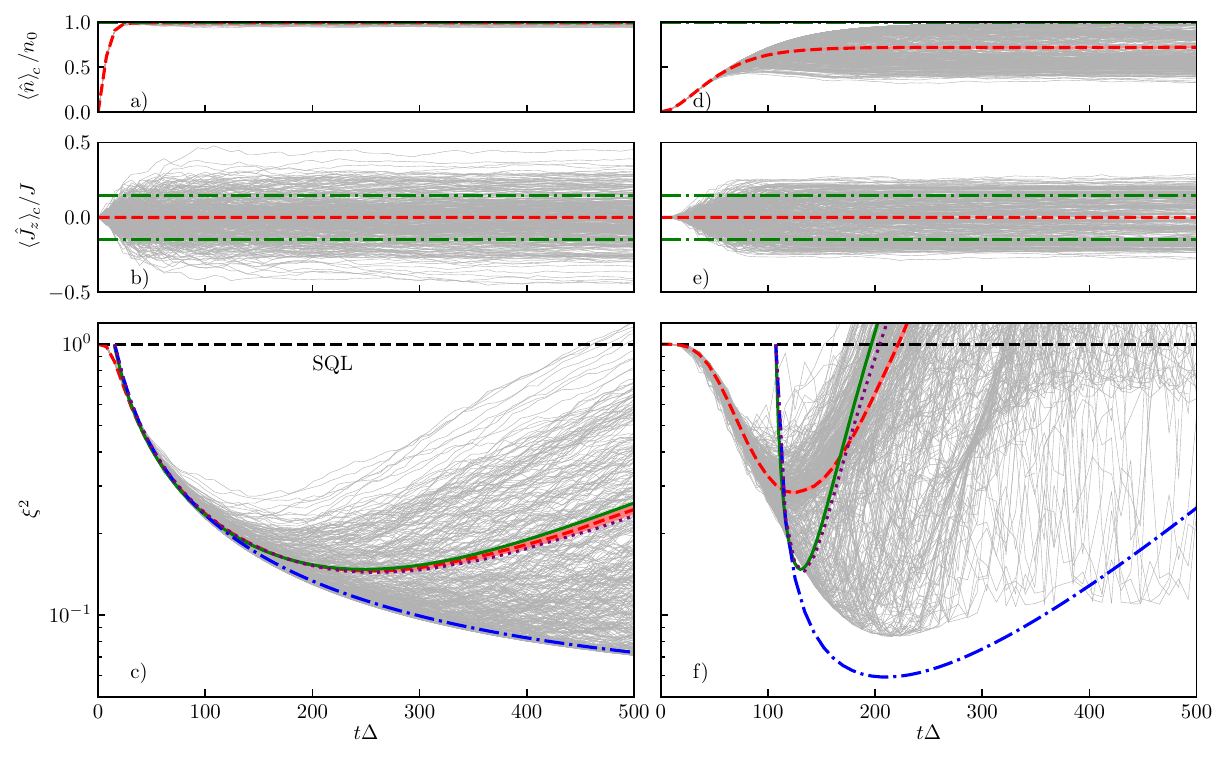}
 \caption{Numerical simulation of the full atom-cavity dynamics in different dissipation regimes, with $N=45$, $g = 0.05\Delta$, and $\eta=1$. Left panels report exemplary trajectories from equation~\eqref{eq:master_equation}, in the bad-cavity regime, with $\kappa = 0.4\Delta, \varepsilon=0.4\Delta$ and $g^2N/\Delta\simeq 0.3 \kappa$.
 Right panels are examples from out of the bad-cavity regime, with $\kappa = 0.04\Delta, \varepsilon = 0.04\Delta$ and $g^2N/\Delta\simeq 3 \kappa$.
 Panels a and d: Distribution of $\langle \hat{n} (t)\rangle_c$ in units of the non-interacting stationary value $n_0$.
 The (red) dashed line is the statistical average $\mathrm{E}[\hat{n} (t)]$. Panels b and e: Distribution of $\langle \hat{J}_z (t)\rangle_c$.
 The (red) dashed line is the statistical average $\mathrm{E}[{J}_z(t)]$, and the (green) dot-dashed lines correspond to the standard deviation of $\hat{J}_z$ in the initial CSS state. Panels c and f: Distribution of $\xi^2(t)$. The (red) dashed line with a band reports the statistical average $\mathrm{E}[\xi^2(t)]$ with its statistical uncertainty from 400 trajectories. The horizontal dashed line is the SQL. The (purple) dotted line is the average $\mathrm{E}[\xi^2(t)]$ for the corresponding adiabatically removed cavity simulations of equations~\eqref{eq:master_equation_cavrem} and the (green) solid curve is the analytical expression~\eqref{eq:analyticalSqueezingAverage}, while the (blue) dot-dashed line is equation~\eqref{eq:fb_squeezing_sol}. 
 Both curves have been shifted by the time at which the number of photons in the full simulations reaches $90\%$ of the stationary value, as extracted from panels a and d.}
 \label{fig:trajectories}
\end{figure*}

\subsection{Numerical results for the full atom-field dynamics}

Having analytically and numerically solved the dynamics in the bad cavity regime in cavity removal approximation, we now focus on the numerical solution of the full SME~\eqref{eq:master_equation}. We are interested in inspecting the accuracy of our previous results and to investigate the main qualitative and quantitative changes to be expected when gradually exiting the bad-cavity regime.

\begin{figure}[tbp]
 \centering
 \includegraphics[width=0.7\columnwidth]{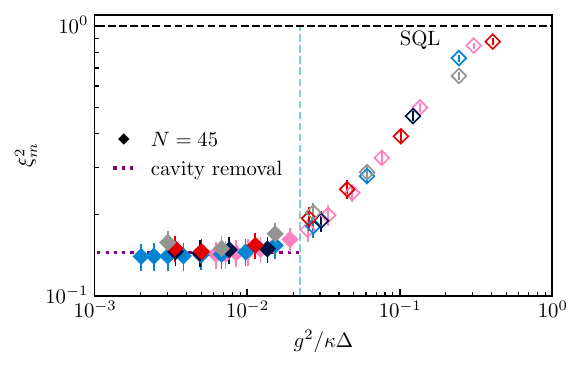}
 \caption{Dependence of $\xi^2_\text{m}$ on $g^2/\kappa\Delta$, for various combinations of parameters (See table~\ref{tab:data_g}). 
 The vertical dashed line indicates the boundary of the bad-cavity condition as defined by~\eqref{eq:bc_condition}, and empty symbols are outside said regime. The dotted line reports the numerical cavity removal result, which, for $N=45$, is more accurate than equation~\eqref{eq:analyticalXi2}.
} 
 \label{fig:xi_dependence_g}
\end{figure}

\emph{Evolution of observables for different trajectories}. In figure~\ref{fig:trajectories}, we show the distribution of relevant observables along different conditional trajectories from the full simulations in the bad-cavity (left panels) and out of the bad-cavity regimes (right panels). We focus on the number of photons in the cavity $\langle \hat{n} (t)\rangle_c$ (panels a and d), the clock population difference $\langle\hat{J}_z (t)\rangle_c$ (panels b and e), and the spin-squeezing parameter $\xi^2$ defined in~\eqref{eq:squeezing_xi_def} (panels c and f).  

In the bad cavity regime, the photon number almost deterministically fills the cavity (panel a), reaching a value close to the non-interacting case $n_0$. It takes a transient time $\delta t = c/(\kappa/2)$ with $c\approx 3$ for the statistical average $\mathrm{E}[\hat{n} (t)]$ to reach $90\%$ of the stationary value.  
During such transient, the population differences (panel b) depart from zero and vary widely, each trajectory tending to fluctuate around a particular eigenvalue of $\hat{J}_z$, since continuous measurement increases the precision in its knowledge.
For small times, most of the trajectories of the spin-squeezing parameter (panel c) are compatible with each other, while, as time progresses, the distribution of $\xi^2(t)$ widens. Some of the trajectories continue to decrease and follow what appears to be an optimal value. This limiting values indeed compare well with the analytical approximate expression in the case of feedback (dot-dashed line), equation~\eqref{eq:fb_squeezing_sol}, provided a temporal shift equal to $\delta t$ is introduced. On the other hand, most of the trajectories tend to increase after reaching a minimum value at some time. Their statistical average also manifests a steep initial decrease, followed by a minimum and a slow increase, until metrological advantage is lost. Hence the reason for characterizing each considered system with the minimum of the average spin-squeezing parameter, $\xi^2_\text{m} = \min_t \mathrm{E}[\xi^2(t)]$. In panel c, we also plot $\mathrm{E}[\xi^2(t)]$ for the adiabatically removed cavity simulation (dotted line), and from the analytical expression~\eqref{eq:analyticalSqueezingAverage} (solid curve); we observe that they are essentially the same as for the full system, provided the initial offset $\delta t$ is introduced.
This indicates that the squeezing process begins as soon as the phase shift induced by the atoms on the photons is detected by the continuous measurement, but the generation rate of spin squeezing reaches its optimal value only when $n(t)$ reaches its stationary value.
On the other hand, the adiabatically removed cavity approximation assumes a stationary photon population, so that the information on the atomic ensemble is directly transmitted to the homodyne detector and spin squeezing is generated right away.
Notice that at short times most of the trajectories of our no-feedback protocol for the full system are compatible with the offset cavity removal results, either excluding (solid or dotted lines) or including (dot-dashed line) feedback. The feedback protocol would require to continuously employ the information gained from the continuous measurement outcomes to realign the atomic state to the equator of the collective Bloch sphere, and guarantee that $\langle\hat{J}_z\rangle_c\approx 0$.
However, as the trajectories evolve and the average collective spins move away from the equator, the tangential planes, on which the metrological spin squeezing is evaluated, are typically less and less parallel to the fixed measurement direction $z$: this causes the departure from the feedback solution.

Our full simulations allow us to consider also scenarios with smaller $\kappa$, outside the bad-cavity regime. Here, the stationary photon number (panel d) varies strongly for different trajectories and is generically significantly lower than in the non-interacting case. The transient $\delta t$, defined as above, corresponds to $c\approx 2.5$ and is longer, due to smaller $\kappa$. It takes therefore a longer time for the population difference to stabilize (panel e), and correspondingly, the dynamics of spin-squeezing generation is strongly mixed with the cavity filling, resulting in much larger variance of the distribution of $\xi^2(t)$ (panel f), even at short times. This has two consequences: first, the minimal average spin-squeezing parameter is worse than in the bad-cavity regime, since most of its trajectories stop decreasing earlier, resulting in a smaller optimal time; second, the cavity removal simulations, with or without feedback, do not provide accurate information on the full system, even introducing a time offset as above.   

\begin{figure}[tb]
 \centering
 \includegraphics[width=0.7\columnwidth]{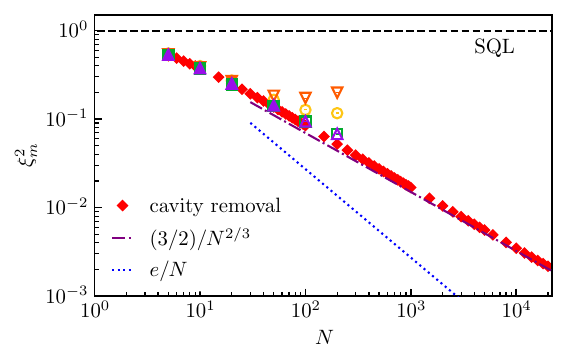}
 \caption{Dependence of $\xi^2_\text{m}$ on $N$, for various combinations of parameters, with $\eta=1$.
 Full (empty) symbols correspond to full simulations in (out of) the bad cavity regime (See table~\ref{tab:data_N}). Diamonds are simulations in the cavity removal approximation.
 The horizontal dashed line indicates the SQL.  The (purple) dot-dashed line is the analytical scaling $\xi^2_{\text{NF,m}}$~\eqref{eq:analyticalXi2}, while the (blue) dotted line is the analytical Heisenberg scaling in presence of feedback $\xi^2_\text{F,m}$~\eqref{eq:fb_squeezing_min}.
 }
 \label{fig:xi_dependence_N}
\end{figure}

\emph{Dependence of $\xi^2_\text{m}$ on coupling at fixed $N$.} Having shown two representative cases, we now consider the dependence of the minimum average spin squeezing in the full system case on the ratio of the effective interaction frequency of the cavity with the atoms $g^2/\Delta$ and the transmission rate $\kappa$. In figure~\ref{fig:xi_dependence_g} we report the $N=45$ case. 
For small values of this ratio, the bad-cavity condition~\eqref{eq:bc_condition} is fulfilled (before the vertical dashed line), and the results converge to the cavity removal simulation with the same number of particles. For this latter case, it is natural for $\xi^2_\text{m}$ to only depend on the number of atoms, since the spin-squeezing parameter is dimensionless and equation~\eqref{eq:master_equation_cavrem} only contains the frequency $\tilde{\kappa}$, which sets the timescale. 
Ratios of other dimensional parameters that only occur in the full master equation \eqref{eq:master_equation}, such as the transmission rate $\kappa$, the driving amplitude $\varepsilon$, and the coupling $g^2/\Delta$, in principle become relevant outside the bad-cavity regime, when also the details of the cavity directly affect the global dynamics. Indeed, here we observe a reduction of spin squeezing with respect to the cavity removal result. Unexpectedly, we still observe small dispersion of the results up to moderate values of $g^2/\kappa\Delta$, the residual variance arguably related to the driving amplitude and thus the stationary number of photons.

\begin{figure}[tbp]
 \centering
 \includegraphics[width=0.7\columnwidth]{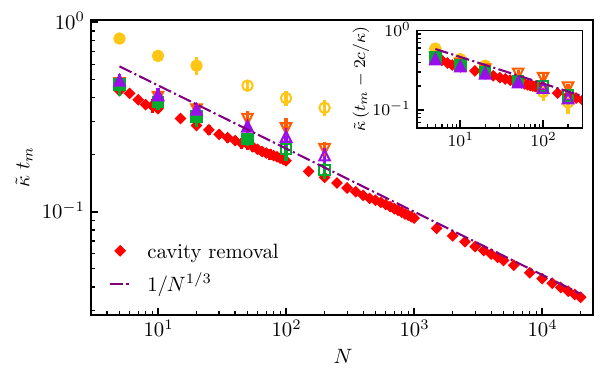}
 \caption{Scaling dependence of optimal time $t_\text{m}$ on the number of atoms $N$, for various combinations of parameters, with $\eta=1$. Full (empty) symbols correspond to full simulations in (out of) the bad cavity regime (See table~\ref{tab:data_N}). Diamonds are simulations in the cavity removal approximation. 
 Main figure shows the optimal time scaled with the corresponding effective rate $\tilde{\kappa}$, estimated using~\eqref{eq:k_eff}.
 Inset shows the optimal time considering the offset introduced in equation~\eqref{eq:t_opt_expr}.
 The dot-dashed lines report the analytical scaling $t_{\text{NF,m}}$ from~\eqref{eq:analyticalTopt}.}
 \label{fig:topt_dependence_N}
\end{figure}

\emph{Scaling of $\xi^2_\text{m}$ with $N$.}
We now discuss the scaling dependence of the spin-squeezing parameter on the atomic ensemble size. In figure~\ref{fig:xi_dependence_N}, we compare the results of the full simulations of different configurations to the numerical results obtained with the adiabatic removal of the cavity, which provide the optimal spin squeezing achievable with this continuous measurement scheme in the absence of feedback. As a reference, we report also the analytical result~\eqref{eq:fb_squeezing_min} from~\cite{Thomsen2002, Thomsen2002a}, for the continuous feedback scheme (dot-dashed line). This reaches the ultimate Heisenberg scaling.
The efficiency of the cavity removal simulations of equation~\eqref{eq:master_equation_cavrem} allows us to consider up to $N=20000$ atoms (diamond symbols).  We then fit the power-law $\xi^2_\text{m}=a/N^\alpha$ and observe convergence in the results, provided only numbers $N\ge 10^3$ are considered. We obtain $a= 1.89(6)$ and $\alpha = 0.680(6)$, which favorably compare to the analytical result~\eqref{eq:analyticalXi2} (dashed line), even though finite-size effects are noticeable.
We simulate the full master equation~\eqref{eq:master_equation} up to $N=200$, for various configurations of $g^2/\Delta$, $\kappa$, $\varepsilon$. 
We confirm the observation that, in the bad-cavity regime (full symbols), $\xi^2_\text{m}$ is independent of any parameter other than $N$, as the results for different configurations are all compatible with each other. 
As the system size increases past $N\gtrsim \kappa\Delta/g^2$ (empty symbols), the results progressively start to deviate from the optimal scaling, even beginning to increase and eventually losing metrological advantage.

\emph{Scaling of the optimal time on $N$.}
In figure~\ref{fig:topt_dependence_N}, we now investigate whether a power law dependence in $N$ holds for the optimal time $t_\text{m}$, for different configurations.
Since in the bad cavity regime the effective dynamics is governed by the effective transmission rate~\eqref{eq:k_eff}, we scale $t_\text{m}$ for each configuration with the corresponding $1/\tilde{\kappa}$.
We then fit the cavity removal results with the power-law $t_\text{m}=b/(\tilde{\kappa}N^\beta)$, obtaining, when considering $N\ge 10^3$, $b=0.9(1)$ and $\beta=0.32(1)$, which are in agreement with the analytical result~\eqref{eq:analyticalTopt}. Concerning the results from the full simulations, we notice that they are mostly compatible with each other and with the cavity-removal ones, once scaled with $\tilde{\kappa}$. However, discrepancies increase with larger driving amplitude (circles), corresponding to large stationary photon number $\sim n_0$. As we commented when discussing figure~\ref{fig:trajectories}, this increase of the optimal time can be modeled by adding a contribution describing the transient $\delta t$ required for filling the cavity, which is initially empty:
\begin{equation}\label{eq:t_opt_expr}
 t_\text{m} = \frac{b}{\tilde{\kappa}N^\beta} + \frac{2 c}{\kappa}\;,
\end{equation}
where $c\simeq 3$ deep in the bad-cavity regime. By removing such transient contribution (inset of figure~\ref{fig:topt_dependence_N}), we indeed observe good agreement among all data. 

\begin{figure}[tbp]
 \centering
 \includegraphics[width=0.7\columnwidth]{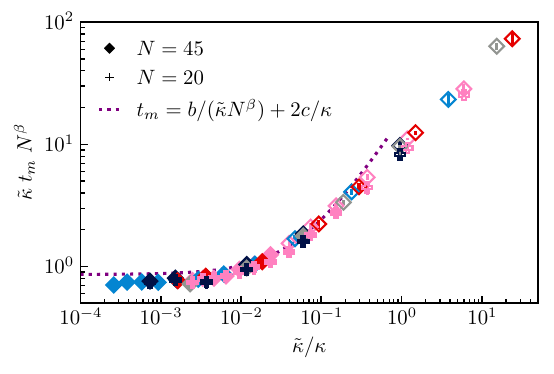}
 \caption{Scaled optimal time dependence on $\tilde{\kappa} /\kappa$ for the full dynamics, for various combinations
of parameters (See table~\ref{tab:data_g}).
 Circles: results for $N = 20$ atoms. Triangles: results for $N = 45$ atoms.
 The dotted line corresponds to~\eqref{eq:t_opt_expr}, with parameters fitted from the cavity removal simulations and including the cavity filling offset.}
 \label{fig:topt_dependence_k}
\end{figure}

\emph{Dependence of $t_\text{m}$ on coupling.}
In figure~\ref{fig:topt_dependence_k}, we focus on the role of coupling in the full simulations in determining the optimal time.
We scale the latter also with the obtained power-law dependence on the atomic ensemble size, and compare two system sizes, $N=20$ and $N=45$.
Unlike for the spin-squeezing parameter in  figure~\ref{fig:xi_dependence_g}, here we notice that a good scaling variable is the ratio of the effective transition rate $\tilde{\kappa}$ with the original transmission rate $\kappa$. This hints at a prominent role of the number of stationary photons, since $\tilde{\kappa}/\kappa=16(g^2/\kappa\Delta)^2 n_0$. A qualitative explanation of figure~\ref{fig:topt_dependence_k} is the following: the squeezing process begins to be considerable only after the cavity has reached the steady state. When $\kappa\gg\tilde{\kappa}$, this filling transient is negligible compared to the squeezing characteristic time, and the bad-cavity adiabatic removal prediction is accurate.
As the ratio $\tilde{\kappa}\,/\,\kappa$ increases, the transient time cannot be neglected, but becomes more and more relevant. As the two time-scales become comparable, $\tilde{\kappa}\,/\,\kappa \sim 1$, once the cavity steady-state is reached the atomic degrees of freedom are already partially squeezed, therefore it takes less to achieve the optimal average spin squeezing than as estimated with~\eqref{eq:t_opt_expr}.

\section{Relevance to V-level optical clocks}\label{sec:SrRel}

Up to this point we have focused on the $\Lambda-$configuration, where a single excited state is coupled to two ground states: this configuration is relevant to describe alkali atoms such as Rubidium where the clock frequency is in the microwave range~\cite{SSMITH_2010,LEROUX_2011}.
However, the same basic scheme can be adapted also to $V-$level atoms (as depicted in figure~\ref{fig:scheme}c), in which a single ground state is coupled to two different excited states. This configuration is relevant for the low-lying levels of alkaline earth-like atoms, such as Strontium, which is the atomic species employed in the cavity-enhanced atomic clock being developed ad INRiM~\cite{TARALLO_2020}.
In this case, the cavity-aided continuous measurement protocol detailed above could be adapted to operate in the proximity of the closed ${}^1S_0 - {}^3P_1$ intercombination transition, which is particularly suitable for continuous measurements because of its extremely low spontaneous emission rate.
Since the ${}^1S_0 - {}^{3}P_0$ clock transition is far-detuned to the cavity and does not directly participate in the dynamics, the $N_\uparrow$ population is constant.
We can still define a collective spin observable as difference of population between the clock states~\cite{Orenes_2022}:
\begin{equation}
 \hat{J}_z = \frac{\hat{N}_\uparrow - \hat{N}_\downarrow}{2} = \frac{N}{2} - \hat{N}_\downarrow
\end{equation}
We stress that this definition is valid under the assumption that the number of atoms on the two clock states $N$ does not change during the dynamics.
Just as for the $\Lambda-$configuration, it is possible to choose a blue detuning $\Delta=\omega_c-\omega_e \gg g_\downarrow\equiv g$ so that the excited state of the cavity-coupled transition can be adiabatically removed (See~\ref{app:elimination}).
In this configuration, the effective Hamiltonian couples the cavity only to the ground state projection operator:
\begin{equation}\label{eq:Sr_hamiltonian}
 \hat{H}_a =
 \sum_{i=1}^{N}\frac{g^2}{\Delta}\,\hat{n}\;\ket{\downarrow}_i\bra{\downarrow}
 = \frac{g^2}{\Delta}\,\hat{n} \;\hat{N}_\downarrow 
 = -\frac{g^2}{\Delta}\,\hat{n} \hat{J}_z + \frac{g^2}{2\Delta}\,\hat{n} N \;.
\end{equation}

The above effective Hamiltonian is basically equivalent to equation~\eqref{eq:ham_eff}, except for a factor $-2$ in the coupling and a constant cavity resonance shift~\cite{Orenes_2022}, which can be neglected in the measurement dynamics. Therefore the results of Sec.~\ref{sec:results} can be adapted to the Sr case. The optimal squeezing time $t_\text{m}$ should be minimized, to reduce spontaneous losses due to absorption of cavity probe light. This can be obtained at the border of the bad cavity regime, $g^2 N/\Delta \simeq \kappa$, which fixes the optimal detuning $\Delta = g^2 N/\kappa$. In this regime, the excited state adiabatic elimination condition~\eqref{eq:erem_cond} requires that the stationary cavity photon number
\begin{equation}
n_0 \ll \left(\frac{g N}{\kappa}\right)^2 \equiv n_{\text{lim}}
\end{equation}
corresponding to an input power limit $P(n_{\text{lim}})=g^2 N^2 \hbar\omega_D/4\kappa\sim N^2\cdot10^{-3}$~pW according to \eqref{eq:num_photons} and the experimental parameters in~\cite{TARALLO_2020} (namely $g\simeq 2\pi\cdot 7$~kHz, $\kappa=2\pi\cdot 30$~kHz, $\omega_D\simeq 2\pi\cdot 429$~THz). We then introduce an attenuation factor $f = P(n_{\text{lim}})/P=n_{\text{lim}}/n_0$ and the optimal time for squeezing is estimated with
\begin{equation}
t_\text{m}^{\text{Sr}} \simeq\frac{f \kappa }{4 g^2 N^{1/3}}+\frac{6}{\kappa}\;.
\end{equation}
In the case of $N=10^4$ and $f=100$, namely $P = 1$~nW, then $t_\text{m}^{\text{Sr}} \simeq 150\;\mu\text{s}$, which is within the current state-of-the-art for continuous quantum measurements of quantum systems~\cite{Magrini2021,Martin_2020}. The expected optimal average spin squeezing parameter would be $\xi^2_\text{m}\simeq (3/2)/N^{2/3} \simeq 25$~dB. The role of incoherent scattering of cavity photons from the adiabatically removed excited state $\ket{e}$, characterized by the spontaneous emission rate $\gamma$, is estimated of the order of $\Gamma_{\text{sc}} = \gamma n_0 g^2/\Delta^2$~\cite{CHEN_2014,Barberena_Tradeoffsunitarymeasurement_2023}. Comparing the effective collective information rate $\tilde{\kappa}$ with an effective collective decoherence rate then yields $\tilde{\kappa}/(\Gamma_{\text{sc}}/N) \propto N 4g^2/(\kappa\gamma)\equiv N\mathcal{C}_0$, namely the collective cooperativity~\cite{TanjiSuzuki2011}, which must be  $N\mathcal{C}_0\gg 1$. This condition must be fulfilled in any conceived QND-induced spin-squeezing experiment and in particular for the proposed protocol~\cite{TARALLO_2020}.

\section{Conclusions}\label{sec:conclusions}

In this work we analytically and numerically analyze the dynamics of a three-level atom coupled to an optical cavity affected by a continuous measurement of the transmitted cavity field.
We show how this continuous measurement observation scheme consistently generates conditional spin-squeezed states. We analyze in detail the corresponding average spin squeezing in the different regimes characterizing the cavity properties and the strength of the interaction between atoms and the cavity mode.
We demonstrate that, in the bad-cavity regime and cavity removal approximation, the achievable optimal average spin squeezing depends solely on the atomic ensemble size with scaling exponent $\alpha=2/3$; complementarily, the optimal duration of the squeezing operation shortens with exponent $\beta=1/3$ on particle number, and depends on an effective information rate. Out of the cavity-removal approximation, we observe that the first correction to this result equals to the short transient required to fill the cavity. Exiting this regime gradually complicates such simple picture and introduces explicit dependence on the pumping parameters. 
The scaling found does not match the ideal results obtained with a continuous feedback scheme, due to the role of the Bloch sphere curvature, as we demonstrate analytically; nevertheless, it is comparable to the scaling for other squeezing methods (e.g. OAT~\cite{KITAGAWA_1993}) and has the additional advantage of relying on a much simpler experimental configuration that does not require a strict feedback control of the atomic system which would introduce further sources of noise~\cite{Cox2016}. 
We considered the role of non-unity measurement efficiency and atomic scattering of the cavity field from the excited state, at the level of Gaussian approximation in the bad-cavity regime. 
Considering these effects also in the full simulations is a major computational challenge that will be addressed in future works, to investigate how they impact the optimal expected average spin squeezing. Also, optimization of the pump laser detuning will be included to investigate the interplay between continuous measurement and cavity-induced interactions~\cite{Li_CollectiveSpinLightLightMediated_2022,Barberena_Tradeoffsunitarymeasurement_2023,Fuderer_Hybridmethodgenerating_2023}. Finally, it would be useful to compare full simulations with the results from the cumulant expansion~\cite{Zhang_StochasticMeanfieldTheory_2023,Plankensteiner_QuantumCumulantsjlJulia_2022,Verstraelen_QuantumClassicalCorrelations_2023} and investigate whether an analytical approach can be pursued also in this case.

The data that support the findings of this study are openly available at~\cite{caprotti_zenodo_2024}.

\ack

This work was supported in part by the European Union’
Horizon 2020 Research and Innovation Program and the EMPIR Participating States through the project EMPIR 17FUN03-USOQS. We acknowledge funding from the QuantERA project Q-Clocks, and from Italian Ministry of Research and Next Generation EU via the PRIN 2022 project CONTRABASS (contract n.2022KB2JJM). We acknowledge the CINECA award IsCb5-PSIOQUAS under the ISCRA initiative, for the availability of high-performance computing resources and support.

\appendix

\section{Adiabatic elimination of the atomic excited level}\label{app:elimination}

The interaction between the single-mode cavity photon field and the ensemble of $N$ three-level uncorrelated atoms, as depicted in figure~\ref{fig:scheme}, panels b and c, is described by the Tavis-Cummings Hamiltonian~\cite{Jaynes_Comparisonquantumsemiclassical_1963,Tavis_ExactSolutionMolecule_1968} extended to two atomic transitions.
Each atom contributes a three-level single-mode Jaynes-Cummings term:
 \begin{equation}\label{eq:H_JC}
 \hat{H}_{JC} = \sum_{j=\downarrow,\uparrow,e}\omega_j \op{j}{j} + g_\downarrow\hat{c}\op{e}{\downarrow} 
 + g_\uparrow\hat{c}\op{e}{\uparrow} + \text{h.c.}
\end{equation}
where the energy of each each atomic level $\ket{j}$ is $\omega_j$ and in general each transition $j\leftrightarrow e$ has a different coupling strength $g_j$ to the single photonic mode described by the bosonic field operator $\hat{c}$.
The ground-state detunings are defined as the difference between the transition frequencies and the cavity frequency $\omega_c$: $\Delta_j = \omega_c - (\omega_e - \omega_j)$, with $j = \downarrow,\uparrow$. The frequency splitting $\omega_\uparrow-\omega_\downarrow = \omega_0$ is the reference clock frequency. We assume that both the detunings and the couplings are uniform across the system.

Based on the coupling strengths $g_j$, the system assumes one of the possible three-level configurations: for example, by taking $g_\uparrow = 0$ we obtain a description for a $V-$level configuration as shown in figure~\ref{fig:scheme}c, typical of alkali-earth atoms such as Sr, in which only the ground state $\downarrow$ is coupled to the cavity mode, whose frequency is of magnitude similar to the clock frequency (see Sec.~\ref{sec:SrRel}).
In this paper, we mainly consider the $\Lambda-$level scheme (see figure~\ref{fig:scheme}b) in which both the $\downarrow$ and the $\uparrow$ levels are coupled to the excited state $e$ via the cavity mode. This configuration is typical of alkali atoms such as Rb, for which $\omega_0\ll\omega_e-\omega_\uparrow\simeq \omega_c$.

It is convenient to perform the transformation of~\eqref{eq:H_JC} to the rotating frame defined by the bare atomic and photonic energies:
\begin{equation}\label{eq:H_rotatingframe}
 \hat{H}_{R} =g_\downarrow e^{-i\Delta_\downarrow t}\hat{c}\op{e}{\downarrow} 
 + g_\uparrow e^{-i\Delta_\uparrow t}\hat{c}\op{e}{\uparrow} + \text{h.c.} 
\end{equation}

If the cavity mode is far-detuned from both the atomic transitions, with respect to the couplings $|g_\uparrow|,|g_\downarrow|\ll|\Delta_\uparrow|,|\Delta_\downarrow|$, the excited state, if initially empty, remains very little populated at the time scales of interest. Therefore it can be adiabatically removed, in order to simplify the interaction which describes the system dynamics.
We briefly recap the time-averaging technique~\cite{GAMEL_2010,JAMES_2011} that allows to perform such removal. Equation~\eqref{eq:H_rotatingframe} is of the harmonic form $\hat{H}=\sum_n \hat{h}_n e^{-i\omega_n t}+h.c.$ which can be approximated by $\hat{H}_{\text{eff}}=\sum_{m,n} [\hat{h}^\dagger_m,\hat{h}_n] e^{i(\omega_m-\omega_n) t}/\omega^+_{mn}+h.c.$, with $(\omega^+_{mn})^{-1}=(\omega_{m})^{-1}+(\omega_{n})^{-1}$, provided $|\omega_m+\omega_n|\gg|\omega_m-\omega_n|$. In our case it is therefore convenient to choose $\hat{h}_1=g_\uparrow \hat{c}\op{e}{\uparrow}$, with $\omega_1=\Delta_\uparrow$, and $\hat{h}_2=g^*_\downarrow \hat{c}^\dagger\op{\downarrow}{e}$, with $\omega_2=-\Delta_\downarrow$, resulting in 
\begin{equation}\label{eq:H_effective}
 \hat{H}_{\text{eff}}=2\left(\frac{|g_\uparrow|^2}{2\Delta_\uparrow}-\frac{|g_\downarrow|^2}{2\Delta_\downarrow}\right)\hat{c}^\dagger\hat{c}\;\hat{s}_z\\+\left(\frac{|g_\uparrow|^2}{2\Delta_\uparrow}+\frac{|g_\downarrow|^2}{2\Delta_\downarrow}\right)\left[\hat{c}^\dagger\hat{c}-(2+3\hat{c}^\dagger\hat{c})\op{e}{e}\right]\;.
\end{equation}

When both the couplings are different from zero, such as in the $\Lambda-$level case, it is convenient to tune the cavity so that $\Delta_\downarrow=-\Delta_\uparrow |g_\downarrow|^2/|g_\uparrow|^2$ and the second term of~\eqref{eq:H_effective} vanishes, resulting in $\hat{H}_{\text{eff}}=2(|g_\uparrow|^2/\Delta_\uparrow)\hat{c}^\dagger\hat{c}\;\hat{s}_z$.
Summing this equation over the atoms yields equation~\eqref{eq:ham_eff}, where we used, without lack of generality, the simplification that the couplings are real and equal, $g=g_\downarrow=g_\uparrow$, and thus $\Delta_{\uparrow} = - \Delta_{\downarrow} = \omega_0/2 \equiv \Delta$.

If instead we consider the $V-$level case with $g_\uparrow=0$, then~\eqref{eq:H_effective} reduces to $\hat{H}_{\text{eff}}=-({|g_\downarrow|^2}/{\Delta_\downarrow})\hat{c}^\dagger\hat{c}\;\hat{s}_z+({|g_\downarrow|^2}/{2\Delta_\downarrow})[\hat{c}^\dagger\hat{c}-(2+3\hat{c}^\dagger\hat{c})\op{e}{e}]$. Summing this equation over the atoms yields equation~\eqref{eq:Sr_hamiltonian}, provided the occupation of the excited state is neglected.

\section{Details of the spin-squeezing parameter estimation}\label{app:rotation}

The evaluation of the conditional spin-squeezing parameter requires the estimation 
of the collective spin components' averages and covariance matrix, as defined in equation~\eqref{eq:cov_mat_def}.
The covariance matrix contains information regarding the variance of the spin components; the optimal spin-squeezing parameter is defined as the variance in the optimal direction on the tangent plane, perpendicular to the average spin $\langle \hat{\vect{J}}\rangle_c = \left(\langle \hat{J}_x\rangle_c, \langle \hat{J}_y\rangle_c, \langle \hat{J}_z\rangle_c\right)$, normalized to the magnitude of such average spin. 
In the simulations, the expectation values $\langle \hat{J}_i (t) \rangle_c$, and the covariance matrix, are referred to the fixed reference system integral to the initial average spin vector along the $x$ direction. In post-processing, therefore, the reference frame at each time step should be passively rotated to the instantaneous average spin vector, after which the relevant covariances appear in the new $y-z$ plane.
This operation is equivalent to the more efficient active rotation of the average spin vector to the $x$ direction, and the corresponding rotation of the covariance matrix. 
The rotation matrix $\mathcal{R}$ necessary to perform such operation in the Euclidean space $\mathbb{R}^3$ is related to the rotation operator $\hat{R}$ in the collective Hilbert space which transforms the spin-coherent state $\ket{\theta, \phi}$ on the Bloch sphere to the initial state $\ket{\frac{\pi}{2},0}$: 
\begin{equation}
 \left.\mathcal{R}_{ij} \langle \hat{J}_{j} \rangle_c \right|_{\theta,\phi}
 = \left.\langle \hat{J}_i \rangle_c \right|_{\pi/2,0}
 = \left.\langle\hat{R}^{-1} \hat{J}_i \hat{R}\rangle_c\right|_{\theta,\phi} \label{eq:rotation_equivalence} 
\end{equation}
from which one gets $\mathcal{R}_{ij} \hat{J}_j = \hat{R}^{-1}\hat{J}_i \hat{R}$ (repeated indexes are summed). The direction $(\theta,\phi)$ is related to the mean spin vector by equations~\eqref{eq:thetaphi} and the rotation operator is a composition of a rotation around the $z$ and $y$ axes:
\begin{equation}
 \hat{R} = \hat{R}_y\left(\tfrac{\pi}{2}-\theta \right) \hat{R}_z(-\phi) = e^{-i \left( {\pi}/{2}-\theta \right) \hat{J}_y} e^{i \phi \hat{J}_z}
\end{equation}
This operator allows to derive the proper transformation of the covariance matrix as
\begin{equation}\label{eq:rotate-covariance}
\left.\text{cov}_{ij}(\hat{\vect{J}}^\prime)\right|_{\pi/2,0}=\left.\mathcal{R}_{ik}\text{cov}_{kl}(\hat{\vect{J}})\mathcal{R}_{jl}\right|_{\theta,\phi}\;.
\end{equation}

Once the covariance matrix is rotated to the fixed reference frame, we can then reduce it to the tangent components $y-z$ and calculate the minimal eigenvalue, from which we obtain the spin-squeezing parameter.

\section{Analytical solution for the spin-squeezing parameter from the master equation in the cavity removal approximation without feedback}\label{app:analytical}

In this Appendix, we report the derivation of the tangential spin-squeezing parameter in the cavity removal approximation in the absence of feedback.

In the cavity removal approximation described by equations~\eqref{eq:master_equation_cavrem}, the mean spin vector always lies in the $x-z$ plane, if the initial state is a CSS along $x$, since there is no Hamiltonian term. Then, the minimal variance from~\eqref{eq:rotate-covariance} is on the rotated $z$ direction and is related to the covariances in the original frame by the following equation:
\begin{equation}
 \Delta^2\hat{J}_\perp=\Delta^2\hat{J}_z\sin^2\theta+\Delta^2\hat{J}_x\cos^2\theta-2\text{cov}\left(\hat{J}_z\hat{J}_x\right)\sin\theta \cos\theta\;,
\end{equation}
where $\theta$ is defined by~\eqref{eq:thetaphi}. From this equation, the tangential spin-squeezing parameter is derived as $\xi^2=(2/J)\Delta^2\hat{J}_\perp/\mathcal{C}$.

To determine $\xi^2$, we study the evolution of the mean spin and the covariance components. A generic observable $\hat{A}$, whose conditional expectation value is $\langle\hat{A}\rangle_c=\Tr[\hat{\rho}_c\hat{A}]$, obeys the following equation, derived from the master equation~\eqref{eq:master_equation_cavrem}:
\begin{equation}\label{eq:observableevolution}
    d\langle\hat{A}\rangle_c = \langle\hat{J}_z\hat{A}\hat{J}_z -\hat{J}_z^2\hat{A}/2 -\hat{A}\hat{J}_z^2/2\rangle_c \;d\tau 
    + \sqrt{\eta}\left(\langle \hat{A}\hat{J}_z+\hat{J}_z\hat{A}\rangle_c- 2\langle\hat{J}_z\rangle_c \langle\hat{A}\rangle_c\right)\; dw\;,
\end{equation}
where we introduced the scaled time $\tau\equiv\tilde{\kappa} t$ and stochastic increment $dw\equiv\sqrt{\tilde{\kappa}}\,dW_t$.

The initial condition for the mean spin is $\langle\hat{J}_z(0)\rangle_c=\langle\hat{J}_y(0)\rangle_c=0$, $\langle\hat{J}_x(0)\rangle_c=J$, while the initial variances are $\Delta^2J_z=\Delta^2J_y=J/2$, $\Delta^2J_x=0$. The off-diagonal covariances are initially zero, and we make our first approximation in setting them to zero for every time:
\begin{equation}
 \text{cov}\left(\hat{J}_i\hat{J}_j(\tau)\right)_c=\frac{1}{2}\langle(\hat{J}_i\hat{J}_j+\hat{J}_j\hat{J}_i)(\tau)\rangle_c-\langle\hat{J}_i(\tau)\rangle_c\langle\hat{J}_j(\tau)\rangle_c = 0\;,
\end{equation}
for $i\neq j$. This is equivalent to third order cumulant truncation, namely Gaussian approximation, and will be confirmed by inspection of the simulation results.

The resulting coupled equations follow straightforwardly from the angular momentum commutation relations and are reported here, where the "c" subscript is understood for all expectation values:
\begin{align}
&d\langle\hat{J}_x\rangle &=\phantom{=}&-\frac{1}{2}\langle\hat{J}_x\rangle \;d\tau + 2\sqrt{\eta}\text{cov}\left(\hat{J}_x\hat{J}_z\right) \;dw \approx -\frac{1}{2}\langle\hat{J}_x\rangle \;d\tau \nonumber\\
&d\langle\hat{J}_y\rangle &=\phantom{=}&-\frac{1}{2}\langle\hat{J}_y\rangle \;d\tau + 2\sqrt{\eta}\text{cov}\left(\hat{J}_y\hat{J}_z\right) \;dw \approx -\frac{1}{2}\langle\hat{J}_y\rangle \;d\tau \nonumber\\
&d\langle\hat{J}_z\rangle &=\phantom{=}& 2\sqrt{\eta}\Delta^2\hat{J}_z \;dw \nonumber\\
&d\langle\hat{J}_x^2\rangle &=\phantom{=}&-\left(\langle\hat{J}_x^2\rangle-\langle\hat{J}_y^2\rangle\right) \;d\tau + \sqrt{\eta}\left(\langle \hat{J}_x^2\hat{J}_z+\hat{J}_z\hat{J}_x^2\rangle- 2\langle\hat{J}_z\rangle \langle\hat{J}_x^2\rangle\right) \;dw \nonumber\\
&d\langle\hat{J}_y^2\rangle &=\phantom{=}&\left(\langle\hat{J}_x^2\rangle-\langle\hat{J}_y^2\rangle\right) \;d\tau 
 + \sqrt{\eta}\left(\langle \hat{J}_y^2\hat{J}_z+\hat{J}_z\hat{J}_y^2\rangle- 2\langle\hat{J}_z\rangle \langle\hat{J}_y^2\rangle\right) \;dw \nonumber\\
&d\langle\hat{J}_z^2\rangle &=\phantom{=}& 2\sqrt{\eta}\left(\langle \hat{J}_z^3\rangle- \langle\hat{J}_z\rangle\langle\hat{J}_z^2\rangle\right) \;dw
\end{align}

When considering  the evolution of the variances, we again make use of third order cumulant truncation~\cite{Albarelli_Ultimatelimitsquantum_2017,MolmerMadsen2004,MADSEN_2004}. This consists in a Gaussian approximation, which is expected to be valid in the $J\to \infty$ limit and is later justified by comparison to numerical results. We then obtain the expressions in equations~\eqref{eq:evolvexy}, which take into account the initial conditions, resulting in equations~\eqref{eq:deltaX}-\eqref{eq:deltaY}.
Notice that due to our approximations, these expressions are unconditional. Besides $\Delta^2\hat{J}_x$, which was not present in~\cite{Thomsen2002,Thomsen2002a}, the other expressions reduce to those in the literature in the small time limit~\cite{Albarelli_Ultimatelimitsquantum_2017}.

The evolution of $\langle\hat{J}_z\rangle_c$ is completely stochastic, and the statistical distribution of conditional values $P\left(\langle\hat{J}_z(\tau)\rangle_c\right)$ is thus constant in time and equivalent to the initial one. In the large $J$ limit, this can be approximated by equation~\eqref{eq:Zdistribution}, namely a Gaussian with zero mean and variance $\Delta^2 \hat{J}_z(0)=J/2$.
The contrast thus evolves as:
\begin{equation}
 \mathcal{C}(\tau)= \frac{\langle\hat{J}_x(\tau)\rangle^2+\langle\hat{J}_y(\tau)\rangle^2+\langle\hat{J}_z(\tau)\rangle_c^2}{J^2}\approx e^{-\tau}\;,
\end{equation}
since $\Delta^2 \hat{J}_z(0)/J^2\to 0$ in the $J\to \infty$ limit.

Both the equations for $\langle\hat{J}_z\rangle$ and $\langle\hat{J}_z^2\rangle$ evolve only stochastically, and one would presume that also $\Delta^2\hat{J}_z=\langle\hat{J}_z^2\rangle-\langle\hat{J}_z\rangle^2$ evolves stochastically. However, this quantity evolves as
\begin{align}
&d\Delta^2\hat{J}_z &=& d\langle\hat{J}_z^2\rangle -d\left(\langle\hat{J}_z\rangle^2\right)
= 2\sqrt{\eta}\left(\langle \hat{J}_z^3\rangle - \langle\hat{J}_z\rangle\langle\hat{J}_z^2\rangle\right) \;dw -4\sqrt{\eta}\langle\hat{J}_z\rangle \Delta^2\hat{J}_z\;dw \nonumber\\
&&\phantom{=}& - 4\eta \left(\Delta^2\hat{J}_z\right)^2 d\tau \nonumber\\
&&=& 2\sqrt{\eta}\langle\hat{J}_z^3\rangle_C\;dw- 4\eta \left(\Delta^2\hat{J}_z\right)^2 d\tau
\end{align}
where the last term stems from the It\^o calculus rule $dx = A\;d\tau + B\;dw \to d(f[x])= \left(A f'[x] + B^2 f''[x]/2\right)d\tau + B f'[x] \;dw$, where $f$ is a function of $x$. $\langle\hat{J}_z^3\rangle_C = \langle\hat{J}_z^3\rangle -3\langle\hat{J}_z\rangle\langle\hat{J}_z^2\rangle+2\langle\hat{J}_z\rangle^3$ is a third order cumulant that we set to zero in Gaussian approximation. The resulting expression, equation~\eqref{eq:deltaZ}, is consistent, for $\eta=1$, large $J$ and moderate times, with the one following heuristically from an approach analogous to~\cite{Thomsen2002,Thomsen2002a}, where one assumes that the atomic state preserves its minimal uncertainty product in the $y-z$ plane: $\Delta^2\hat{J}_z\Delta^2\hat{J}_y= |\langle \hat{J}_x\rangle|^2/4$, yielding:
\begin{equation}
 \frac{\Delta^2\hat{J}_z(\tau)}{J/2} \approx 
 \frac{e^{-\tau}}{J\left( 1-e^{-2\tau}\right)+ \left( 1+e^{-2\tau}\right)/2}\;.
\end{equation}

We now have all the ingredients for determining  spin-squeezing as a function of time and of the $\langle\hat{J}_z\rangle_c$ projection, recalling its relation to $\cos{\theta}$ in~\eqref{eq:thetaphi}. Neglecting again the off-diagonal covariances, we obtain:
\begin{equation}\label{eq:analyticalSqueezingCosTheta}
 \xi^2(\tau,\langle\hat{J}_z\rangle_c^2)=\frac{e^\tau}{1+2\eta J\tau}\left(1-\langle\hat{J}_z\rangle_c^2/|\langle \hat{\vect{J}}\rangle|^2\right)
 + e^\tau \left[J\left( 1-e^{-\tau}\right)^2+ \frac{1}{2}\left( 1-e^{-2\tau}\right)\right]  \frac{\langle\hat{J}_z\rangle_c^2}{|\langle \hat{\vect{J}}\rangle|^2}\;.
\end{equation}

We use the above expression in the main text to derive the scaling of the average spin-squeezing parameter and to that end perform the $J\to\infty$ limit, in particular for the contrast. However, since this expression should be valid even for large $\langle\hat{J}_z\rangle$, in this case it can be more accurate to retain $\mathcal{C}(\tau)= e^{-\tau} + \langle\hat{J}_z(\tau)\rangle_c^2/{J^2}$. We also notice that performing the average of this more refined expression would introduce corrections in terms of powers of $e^\tau/2J$, which do not affect the found scaling of the minimum.

\section{Relation between scaling exponents of spin squeezing and optimal time}\label{app:exponents}

The relation $\alpha+\beta = 1$ stems from a generic behavior of the spin-squeezing parameter for small times, in which $\xi^2(t) \approx Q^{-\delta} + f Q^\gamma/N^\epsilon$ with $Q\propto N t$. Indeed, the minimum of this function occurs at $\xi^2_\text{m}=(1+\delta/\gamma)(\delta/f\gamma)^{\delta/(\gamma+\delta)}N^{-\delta\epsilon/(\gamma+\delta)}$ for $\bar{Q}=(\delta/f\gamma)^{1/(\gamma+\delta)}N^{\epsilon/(\gamma+\delta)}$, implying $\alpha=\delta\epsilon/(\gamma+\delta)$ and $\beta=1-\epsilon/(\gamma+\delta)$. The relation $\alpha+\beta=1$ is thus valid if $\delta=1$, which holds in our cavity-removal case, as can be seen in figure~\ref{fig:trajectories} in the bad-cavity regime and equation~\eqref{eq:fb_squeezing_sol} for large $N$. Notice that for OAT, although $\alpha=2/3$ as in our model, $\beta=2/3$ instead of $\beta=1/3$, since in that case $\delta=2$ and thus $\alpha+2\beta=2$.

\section{Details of the numerical simulations}\label{app:simulation}

Given the stochasticity of the evolution due to the explicit dependence on the measurement outcome, each solution of equation~\eqref{eq:master_equation} represents a different unraveling, a particular trajectory of the conditional dynamics, namely a model for a specific realization of an experiment. 
Therefore a particular configuration of physical parameters $g, \kappa, \varepsilon, N$ can be generically characterized only based on the behavior of the system averaged over many trajectories.
Each trajectory is found by integrating the conditional master equations, using the \texttt{QuTiP 4.7} library~\cite{QUTIP_2012,QUTIP_2013}.
Since the initial state is pure, and we consider $\eta=1$ in the simulations, we use the \texttt{ssesolve} dynamic solver for the stochastic Schrödinger equation, which needs less computational resources than the equivalent solver \texttt{smesolve} for master equations. 
This solver implements the \emph{implicit Milstein} method, which we found to be the most accurate at long times among those available, at a relatively moderate cost. This method is indeed known for being suited to stiff dynamics, where very different time scales are relevant in the evolution~\cite{Kloeden_NumericalSolutionStochastic_1995}, such as in the full atom-cavity system that we consider.

The QuTiP library offers the possibility to automatically evolve different trajectories in parallel, profiting of multi-core CPUs, and finally yielding the average of the observables.
However, since the metrological spin-squeezing parameter $\xi^2$ is not associated to a single quantum operator, but it is the ratio of expectation values of different operators, it cannot be evaluated directly by the QuTiP library during the evolution and must be evaluated in post-processing from the expectation values of the relevant observables.
Were we to compute the spin-squeezing parameter from the average observables, we would obtain a result corresponding to the unconditional evolution where no continuous measurement is executed and no squeezing is generated.
Therefore, $\xi^2$ must be evaluated specifically for each trajectory, and its average and standard deviation are then statistically estimated.

The number of trajectories $M$ determines the precision of the results, and we estimate that $M=100 \divisionsymbol 1000$ is sufficient for our purposes. 
In order to have both a statistically relevant sample of trajectories but also to speed up the computation we rely on parallel numerical methods, we employ the \verb|parallel_map| tool provided by QuTiP to solve simultaneously different trajectories, each parameterized by independent seeds that initialize the stochastic increments of the SSE.
This implementation is embarrassingly parallel and the speed-up thus grows linearly with the number of available cores.

The computational complexity of the simulations is proportional to the total number of integration steps.
The minimal number of time-steps required to achieve a suitable level of accuracy for the integration of the SSE is determined in order to guarantee the resolution of any process, without accidentally time-averaging any higher-frequency effect. We therefore compare the most relevant frequencies in the master equations, including: the effective atomic shift $n_0\;\delta\omega$, the effective cavity shift $N\;\delta\omega $, the decay rate $\kappa$ and the driving strength $\varepsilon$.
The time step is then chosen as $dt = 2\pi/R \omega_{\text{max}}$, where the number $R=1000$ has been estimated to be sufficient to yield acceptable accuracy, that is compatibility with the true value, extrapolated for $dt\to 0$, at the precision obtained given the chosen number of trajectories $M=1000$. For the minimal average spin-squeezing parameter, we noticed a residual time-step bias that we estimated as $\delta \xi^2_\text{m}\simeq 1.5\cdot 10^{-2}$ and added to the uncertainty bars.

The other major contribution to computational complexity is given by the size of the quantum system: the SME resolution would require the complete density matrix, therefore the memory usage would grow as $(d_c d_a)^2$, where $d_c$ is the dimension of the photonic Hilbert space and $d_a$ is the dimension of the atomic Hilbert space. It is immediately clear that solving the SSE is beneficial because the memory requirement only grows as $d_c d_a$. As customary, the photonic Fock space is cut off at a maximum number of photons that we expect to be relevant in the considered dynamics. Given the predictions of~\eqref{eq:num_photons}, we can estimate the expected number of photons at the steady state from the initial parameter, thus also giving an estimate of the required dimension to avoid a too low cut-off.
Since for coherent photonic states in the uncoupled steady state we would have $\Delta^2 \hat{n} = n_0$, to be more conservative for generic coupled dynamics, we set $d_c = \texttt{int}(3\,n_0+6)$. The atomic Hilbert space dimension in the initial qutrit representation would scale as $d_a=3^N$. This allows to perform simulations with up to $N\simeq 10$ atoms (not shown in this work) with standard resources. The adiabatic removal of the excited state allows for reducing the dimension to $d_a=2^N$, allowing for the simulation of up to $N\simeq 16$. However, not considering atomic scattering gives us the possibility to restrict the dynamics to the atomic Dicke sector with maximum eigenvalue $J(J+1)$ of $\hat{J}^2$, with $J=N/2$, whose space dimension is $d_a=N+1$. This allows us to simulate up to $N=200$ atoms when considering the full SSE corresponding to~\eqref{eq:master_equation}, and $N=20000$ after performing the adiabatic removal of the cavity.

\section{Data series included in the Figures}

\begin{table}[h]
\centering
    \begin{tabular}{||c||c|c||}
        \hline
         Symbol & $\varepsilon/\Delta$ & $\kappa/\Delta$  \\
         \hline
         $\MyDiamond[HotPink1,fill=HotPink1]$
         & 0.002  & 0.002  \\
         $\MyDiamond[DodgerBlue1,fill=DodgerBlue1]$ & 0.01    & 0.01 \\
         $\MyDiamond[Blue3,fill=Blue3]$ & 0.02 & 0.02   \\
         $\MyDiamond[red,fill=red]$ & 0.009 & 0.006 \\
         $\MyDiamond[Ivory4,fill=Ivory4]$ & 0.02 & 0.01 \\
        \hline
    \end{tabular}
    \caption{Simulation parameters for figures~\ref{fig:xi_dependence_g} and~\ref{fig:topt_dependence_k} and corresponding symbols.}\label{tab:data_g}
\end{table}

\begin{table}[h]
\centering
    \begin{tabular}{||c||c|c|c||}
        \hline
         Symbol & $\varepsilon/\Delta$ & $\kappa/\Delta$ & $g/\Delta$  \\
         \hline
         \tikzcircle{2.5pt} & 0.1 & 0.1   & 0.494  \\
         \color{DarkOrange1}{$\blacktriangledown$} & 0.02  & 0.08  & 0.494 \\
         \color{Green3}{$\blacksquare$}& 0.1  & 0.2  & 0.494  \\
         \color{DarkOrchid4}{$\blacktriangle$}& 0.2  & 0.2  & 0.494 \\
        \hline
    \end{tabular}
    \caption{Simulation parameters for figures~\ref{fig:xi_dependence_N} and~\ref{fig:topt_dependence_N} and corresponding symbols.}\label{tab:data_N}
\end{table}

\bibliographystyle{iopart-num}

\providecommand{\newblock}{}

\end{document}